\documentclass[fleqn,usenatbib]{mnras}

\usepackage{newtxtext,newtxmath}
\usepackage{amsmath}
\usepackage{graphicx}
\usepackage{textcomp}

\usepackage[T1]{fontenc}

\DeclareRobustCommand{\VAN}[3]{#2}
\let\VANthebibliography\thebibliography
\def\thebibliography{\DeclareRobustCommand{\VAN}[3]{##3}\VANthebibliography}

\usepackage{graphicx}	
\usepackage{amsmath}	

\title[Stellar Ages of M Dwarf Hosts]{Stellar Ages of M Dwarf Hosts of Temperate Sub-Neptunes}

\author[Sairam et al.]{
Lalitha Sairam,$^{1}$\thanks{ lalitha.sairam@ast.cam.ac.uk}
and Nikku Madhusudhan,$^{1}$\thanks{ nmadhu@ast.cam.ac.uk}
\\
$^{1}$Institute of Astronomy, University of Cambridge, Madingley road, Cambridge CB3 0HA, UK\\
}

\date{Accepted XXX. Received YYY; in original form ZZZ}

\pubyear{\the\year{}}

\begin{document}
\label{firstpage}
\pagerange{\pageref{firstpage}--\pageref{lastpage}}
\maketitle

\begin{abstract}

Measuring M dwarf ages is a central challenge in stellar astrophysics. These stars evolve so slowly that isochrone fitting gives little constraint, requiring alternative methods based on rotation, activity, and kinematics. Reliable ages help constrain the formation and evolution of planetary systems around M dwarfs, including the temperate sub-Neptunes now being characterised with the \textit{James Webb Space Telescope} (JWST). We present an age analysis of six nearby M dwarf planet hosts (TOI-732, TOI-736, TOI-270, TOI-1468, TOI-1231, and K2-18) using lithium absorption at 6708,\AA, rotation periods, and Galactic kinematics. The Li~I 6708,\AA\ feature is absent above 3$\sigma$ in all targets, implying substantial depletion and ages exceeding 200\,Myr. Rotation periods of 39--145 days yield ages of 2.8--8.6\,Gyr, calibrated against open cluster M dwarfs and including the intrinsic dispersion of the rotation--age relation. Kinematic ages span 0.8--13\,Gyr with uncertainties of 4--6\,Gyr. These are less precise per star but help identify outliers such as TOI-1231 ($\sim$13\,Gyr) and K2-18 ( $\sim$1--2\,Gyr) and place the stars in Galactic context. Rotation yields age constraints 2--6 times tighter than kinematics where periods are measured, and comparable constraints where periods must be inferred from chromospheric activity. Combining multiple indicators enables more secure classification of these stars' evolutionary states, providing reference points for future comparative studies.
\end{abstract}





\section{Introduction}

M dwarfs are the most abundant stellar population in the Galaxy, and their ubiquity makes them key targets for exoplanet surveys \citep{Henry_2004, Bastian2010, Reyle_2021}. Recent discoveries of low-mass exoplanets in close-in orbits around M dwarfs, including candidates for Hycean worlds \citep{Madhusudhan_2021}, have renewed interest in understanding the physical and evolutionary properties of these stars. Observations with the JWST have further emphasised the importance of temperate sub-Neptunes around M dwarfs as prime targets for atmospheric characterisation. K2-18b has been observed with NIRSpec and NIRISS, revealing prominent CH$_4$ and CO$_2$ absorption consistent with a hydrogen-rich atmosphere \citep{Madhusudhan2023, Madhusudhan_2025,Hu2025}. Similarly, JWST transmission spectroscopy of TOI-270d with NIRSpec G395H has detected CH$_4$ and CO$_2$ at high significance, consistent with expectations for a candidate Hycean world \citep{Holmberg2024}, while complementary analyses indicate a highly metal-rich atmosphere with a miscible H$_2$/He–volatile envelope \citep{Benneke2024}. These results demonstrate the capability of JWST to probe the atmospheric composition of temperate sub-Neptunes and set the stage for complementary studies of their host stars.

However, one of the primary challenges in characterising M dwarf planetary systems remains the determination of stellar ages, which are essential for constraining the formation and evolution of their planets \citep{Newton2018, Engle2023}. Due to their low masses and slow nuclear burning rates, M dwarfs evolve on timescales that exceed the Hubble time \citep{Laughlin1997, Baraffe_1998, Dotter2008, Choi_2016, Engle2023}. As a result, conventional age-dating techniques, such as isochrone fitting, are largely ineffective for the majority of field M dwarfs. Alternative methods, including rotation, activity, and lithium depletion, have been used to estimate ages, but each of these indicators has limitations, particularly for older or fully convective M dwarfs.

Lithium is a well-established youth indicator for low-mass stars, but in M dwarfs, lithium is depleted on timescales that depend sensitively on mass and internal structure \citep{Jeffries2005, Martin2018}. While lithium absorption can confirm extreme youth, its absence cannot be used to place precise upper limits on age beyond a few hundred Myr. Similarly, magnetic activity tracers, such as H$\alpha$ and X-ray emission, are linked to stellar rotation but show significant scatter at fixed age and mass \citep{West_2008, Shulyak2019}. Fully convective M dwarfs, in particular, maintain strong magnetic fields and high levels of activity well into Gyr timescales, making it difficult to derive robust activity-age relations \citep{Newton2017, Meunier2024}.

Rotation periods offer an additional constraint on age via gyrochronology \citep{Barnes_2007, Mamajek_2008}, but for M dwarfs, spin-down is a complex function of mass, age, and magnetic topology. While early M dwarfs (M0–M3) show signs of angular momentum loss over Gyr timescales, later types remain rapid rotators for much longer \citep{Irwin_2011,  Garraffo_2018}. The calibration of rotation-age relations for fully convective M dwarfs (below $\sim$0.35 $M_\odot$) remains incomplete, with evidence for a wide dispersion in rotation rates at fixed age \citep{Douglas_2017, Reiners_2018, Jeffers_2022}.

Galactic kinematics, including velocity dispersion and position relative to the Galactic plane, provide an additional statistical handle on age \citep{Leggett_1992, Montes_2001, Newton_2016, Almeida2018}. However, kinematic ages are probabilistic in nature and offer limited precision for individual stars, particularly for those younger than $\sim$1 Gyr where the velocity dispersion is low and overlapping across age groups.

Given these challenges, a combination of independent diagnostics is required to obtain reliable age estimates for M dwarf hosts of planets. Accurate stellar ages are essential for evaluating the atmospheric evolution and habitability potential of close-in planets, particularly temperate sub-Neptunes that may be vulnerable to atmospheric escape and photochemical evolution under prolonged high-energy irradiation \citep{Lalitha2014, Loyd_2018, France2020,  Fromont_2024}. 
In this work, we present a systematic analysis of six M dwarfs hosting temperate sub-Neptunes (TOI-736, TOI-270, TOI-1468, TOI-1231, K2-18, and TOI-732) using a combination of lithium absorption, magnetic activity indicators, rotation measurements, and kinematic properties to constrain their ages. Two of these systems, K2-18 and TOI-732, were previously analysed by \citet{Lalitha2025}, where stellar ages were inferred from their measured rotation periods using the same empirical framework adopted here. In the present study, we revisit these systems to provide a unified and self-consistent analysis that combines rotation-based ages with complementary diagnostics from kinematics and stellar activity. By integrating multiple indicators, we aim to build a consistent framework for M dwarf age determination and to assess the implications of stellar ages for the atmospheric and dynamical evolution of planets in these systems.

Section~\ref{sec:targets} presents the target stars and their properties. Section~\ref{sec:analysis} describes the age diagnostics from lithium absorption, rotation, and kinematics. Section~\ref{sec:results_discussion} presents the results, and Section~\ref{sec:summary} summarises the conclusions.

\section{Targets}\label{sec:targets}
Our sample consists of six M dwarf stars hosting temperate sub-Neptunes: TOI-736, TOI-270, TOI-1468, TOI-1231, K2-18, and TOI-732.  These systems were selected as nearby, well-characterised M dwarfs with confirmed temperate sub-Neptunes that are of particular interest for ongoing and upcoming JWST observations. The sample is not intended to be complete but rather representative of the best-characterised systems suitable for a uniform, multi-diagnostic age analysis. Two of these stars, K2-18 and TOI-732, were previously analysed by \citet{Lalitha2025} and are included here to ensure consistency with the present analysis.
Table~\ref{tab:target} summarises the key astrometric, kinematic, and stellar parameters adopted for each target.

\begin{itemize}

\item TOI-736: Also known as LP 791-18, is an M6 dwarf star located at a distance of 26.5 parsecs from Earth (\citealt{Gaia2021}), with an effective temperature of 2980$\pm$50 K (\citealt{Crossfield2019}). The system hosts three confirmed exoplanets: LP 791-18 b, a terrestrial planet with a radius of 1.12$\pm$0.13 R${\oplus}$ (\citealt{Crossfield2019}); LP 791-18 c, a sub-Neptune with a radius of 2.31$\pm$0.25 R${\oplus}$ and a mass of 7.1$\pm$0.7 M${\oplus}$ (\citealt{Crossfield2019}, \citealt{Peterson2023}); and LP 791-18 d, a planet near the inner edge of the habitable zone with a radius of 1.5$\pm$0.1 R${\oplus}$ (\citealt{Peterson2023}). Their orbital periods are 0.95 days, 4.98 days, and 8.73 days, respectively (\citealt{Crossfield2019}, \citealt{Peterson2023}). Previous studies indicate that it is an old field M dwarf, with an estimated age of $\gtrsim0.5$~Gyr and likely several Gyr (\citealt{Crossfield2019}).

\item TOI-270: This system, discovered in 2019, consists of the M3.0V dwarf star TOI-270, located at a distance of 22.4 $\pm$ 0.1 parsecs (\citealt{Gaia2023}). It hosts three confirmed planets: TOI-270 b, a rocky super-Earth with a radius of 1.26 $\pm$ 0.03 R${\oplus}$ and a mass of 2.4 $\pm$ 0.3 M${\oplus}$; TOI-270 c and TOI-270 d, two mini-Neptunes with radii of 2.6 $\pm$ 0.1 R${\oplus}$ and 2.4 $\pm$ 0.1 R${\oplus}$, respectively (\citealt{Gunther2019}). The orbital periods of 3.36 days, 5.70 days, and 11.4 days suggest a near-resonant configuration, which may have influenced their formation and evolution. No stellar age estimate has been reported for this system in the literature.

\item TOI-1468: Discovered in 2022, this M-dwarf star is located $24.720\pm0.036$ parsecs from Earth (\citealt{Gaia2023}). With an effective temperature of 3360 $\pm$ 51 K (\citealt{Chaturvedi2022}), TOI-1468 exhibits low activity, providing a stable environment for its planetary system. The system hosts two confirmed exoplanets, TOI-1468 b and TOI-1468 c (\citealt{Chaturvedi2022}), which are positioned on opposite sides of the radius valley, a region where planets transition from rocky, Earth-like compositions to gaseous, Neptune-like compositions. \citet{Chaturvedi2022} reported that the stellar age is largely unconstrained, with estimates in the range of 1–10 Gyr.

\item TOI-1231: Located 27.6227 $\pm$ 0.0609 pc away, TOI-1231 is an M3 dwarf star with an effective temperature of 3535 $\pm$ 51 K (\citealt{Burt2021}). The system hosts at least one confirmed exoplanet, TOI-1231~b, a Neptune-like planet with a mass of 15.4~$\pm$~3.3~$M_{\oplus}$ and a radius of 3.65~$\pm$~0.16~$R_{\oplus}$, making it comparable in size and only slightly smaller than Neptune \citep{Burt2021}. This planet orbits its host star with a period of 24.2456 $\pm$ 0.0001 days at a distance of 0.1288 $\pm$ 0.002 AU (\citealt{Burt2021}). No stellar age estimate was reported in \citet{Burt2021}.

\item K2-18: An M2.8 dwarf star located approximately 38 parsecs from Earth (\citealt{Gaia2021}), K2-18 has an effective temperature of 3500$\pm$50 K (\citealt{Benneke2019}) and hosts a planetary system of significant interest for atmospheric studies. The system contains two confirmed planets: K2-18 b, a sub-Neptune with a radius of 2.61$\pm$0.09 R${\oplus}$ and a mass of 8.6$\pm$1.3 M${\oplus}$, and K2-18 c, whose properties are less well-constrained (\citealt{Cloutier_2017}). K2-18 b orbits within the habitable zone with a period of 33 days, receiving stellar flux similar to Earth. Its size and density are consistent with a Hycean planet, suggesting a thick hydrogen-rich atmosphere over a water-dominated interior (\citealt{Madhusudhan_2021}). Recent JWST observations have revealed methane and carbon dioxide in the atmosphere of K2-18 b (\citealt{Madhusudhan2023}). \citet{Lalitha2025} derived a stellar age of $2.9-3.1$ Gyr from its measured rotation period.

\item TOI-732: This M3.5 dwarf star, located 22.03 $\pm$ 0.017 parsecs from Earth (\citealt{Gaia2023}), was discovered in 2020. With an effective temperature of 3360 $\pm$ 51 K (\citealt{Nowak2020}), TOI-732 exhibits relatively low activity, providing a stable environment that reduces the risks posed by stellar flares and high-energy radiation. This makes the system an intriguing target for studying the interplay between stellar activity and planetary habitability. It hosts at least two confirmed exoplanets, TOI-732 b and TOI-732 c (\citealt{Cloutier_2020, Nowak2020}). Based on its measured rotation period, \citet{Lalitha2025} estimated the stellar age to be $6.7-8.6$ Gyr.
    
\end{itemize}

\begin{table*}
    \centering
        \setlength{\tabcolsep}{3pt}
        \caption{Stellar parameters for our targets.}
\begin{tabular}{lcccccc}
    \hline
    Target & TOI-736 & TOI-270 & TOI-1468 & TOI-1231 & K2-18 & TOI-732 \\
    Parameters &  &  &  &  &  &  \\
    \hline
    \hline
    $\alpha$ [deg] & $165.691\pm0.036$ & $68.415\pm0.014$ & $16.654\pm0.029$ & $159.747\pm0.013$ & $172.560\pm0.020$ & $154.646\pm0.020$ \\
    $\delta$ [deg] & $-16.406\pm0.029$ & $-51.956\pm0.013$ & $19.225\pm0.024$ & $-52.469\pm0.012$ & $7.588\pm0.021$ & $-11.716\pm0.021$ \\
    $\mu_{\alpha}$ [mas yr$^{-1}$] & $-221.291\pm0.043$ & $83.082\pm0.017$ & $-42.067\pm0.047$ & $-89.394\pm0.019$ & $-80.479\pm0.026$ & $-341.537\pm0.032$ \\
    $\mu_{\delta}$ [mas yr$^{-1}$] & $-58.841\pm0.038$ & $-268.803\pm0.018$ & $-222.790\pm0.032$ & $361.546\pm0.015$ & $-133.007\pm0.023$ & $-247.747\pm0.032$ \\
    Distance [pc] & $26.652\pm0.028$ & $22.477\pm0.014$ & $24.720\pm0.036$ & $27.480\pm0.016$ & $38.009\pm0.031$ & $22.027\pm0.031$ \\
    $v_{\mathrm{r}}$ [km s$^{-1}$] & $14.100\pm0.100^a$ & $25.90\pm0.370$ & $10.851\pm0.260$ & $70.501\pm0.014$ & $0.328\pm0.014$ & $0.271\pm0.341^h$ \\
    U [km s$^{-1}$] & $-20.522\pm0.301$ & $21.607\pm0.396$ & $8.540\pm0.331$ & $-20.069\pm1.126$ & $-1.398\pm0.014$ & $-15.087\pm0.340$ \\
    V [km s$^{-1}$] & $-23.664\pm0.281$ & $-32.438\pm0.143$ & $-6.430\pm0.205$ & $-73.373\pm1.126$ & $-25.151\pm0.004$ & $-22.575\pm0.004$ \\
    W [km s$^{-1}$] & $-7.381\pm0.283$ & $-7.520\pm0.141$ & $-26.635\pm0.209$ & $39.215\pm1.127$ & $-12.407\pm0.004$ & $-34.699\pm0.005$ \\
    V [mag] & $16.910\pm0.200$ & $12.603\pm0.03$ & $12.500\pm0.200$&$12.360\pm0.200$ & $13.477\pm0.042$& $13.140\pm0.035$\\
    K [mag]& $10.644\pm0.023$ & $8.251\pm0.02$ & $8.497\pm0.017$& $8.069\pm0.026$& $8.899\pm0.019$& $8.204\pm0.021$ \\
    $\log_{10}$(L/L$_{\mathrm{bol}}$) & $0.061\pm0.011$ & $0.252\pm0.081$ & $-0.028\pm0.047$& $0.010\pm0.006$& $0.131\pm0.066$ &$0.052\pm0.011$ \\
    T$_{\mathrm{eff}}$ [K] & $2960\pm55^a$ & $3506\pm70^c$ & $3496\pm25^d$ & $3553^{+51}_{-52}$$^e$ & $3645\pm52^f$& $3358\pm92$$^f$\\
    Spectral Type & M6V$^a$ & M3V$^c$ & M3V$^d$ & M3V$^e$ & M2.5V$^g$ & M4V$^i$ \\
    Fe/H [dex] & $-0.09\pm0.19^a$ & $-0.20\pm0.12^c$& $-0.04\pm0.07^d$ & $0.041^{+0.069}_{-0.063}$$^e$& $0.10\pm0.12^f$& $0.22\pm0.13$$^f$\\
    Radius [R$_{\odot}$] & $0.182\pm0.007^b$ & $0.378\pm0.011^c$& $0.344\pm0.005^d$ & $0.476^{+0.015}_{-0.014}$$^e$ & $0.496\pm0.010^g$&$0.380\pm0.012$$^h$  \\
    Mass [M$_{\odot}$] & $0.139\pm0.005^a$ & $0.386\pm0.008^c$& $0.339\pm0.011$$^d$ & $0.485\pm0.024$$^e$& $0.495\pm0.004^g$&$0.381^{+0.024}_{-0.034}$$^h$ \\
    $v \sin i$ [km s$^{-1}$] & $<2^a$ & --& $<2$~$^d$ &-- & --& $\leq$ 3.4$^h$\\
    \hline
\end{tabular}
    
Notes: a) \cite{Crossfield2019},  b) \cite{Peterson2023}, c) \cite{Van_Eylen_2021}, d) \cite{Chaturvedi2022}, e) \cite{Burt2021}, f) \cite{Lalitha2025}  g) \cite{Cloutier_2017} h) \cite{Cloutier2019} i)  \cite{Bonfanti2024}. V and K magnitudes are from TIC v8.2 and 2MASS, respectively. UVW velocities for all the targets  are calculated in this work.
    \label{tab:target}
\end{table*}

\section{Analysis} \label{sec:analysis}

In this section, we present age estimates for the target stars based on independent diagnostics, which together provide constraints on their evolutionary status.

\subsection{Age constraints from  Lithium abundance}

We conducted a uniform analysis of the Li\,\textsc{i}\,6708\,\AA\ feature for all targets using the highest signal-to-noise (SNR) optical spectrum available in each case.  
The data were obtained from the public archives of ESPRESSO (\(R\simeq140{,}000\)), HARPS (\(R\simeq115{,}000\)), and CARMENES-VIS (\(R\simeq94{,}600\)).  
Each spectrum was corrected to the stellar rest frame and locally continuum-normalised over 6702–6714\,\AA.  
The per-pixel SNR near 6708\,\AA\ was derived consistently across all instruments using the same line-free continuum bands (6702–6706\,\AA\ and 6710–6714\,\AA), as described in Appendix~\ref{app:snr_calc}.

\begin{figure}
\includegraphics[width=0.45\textwidth]{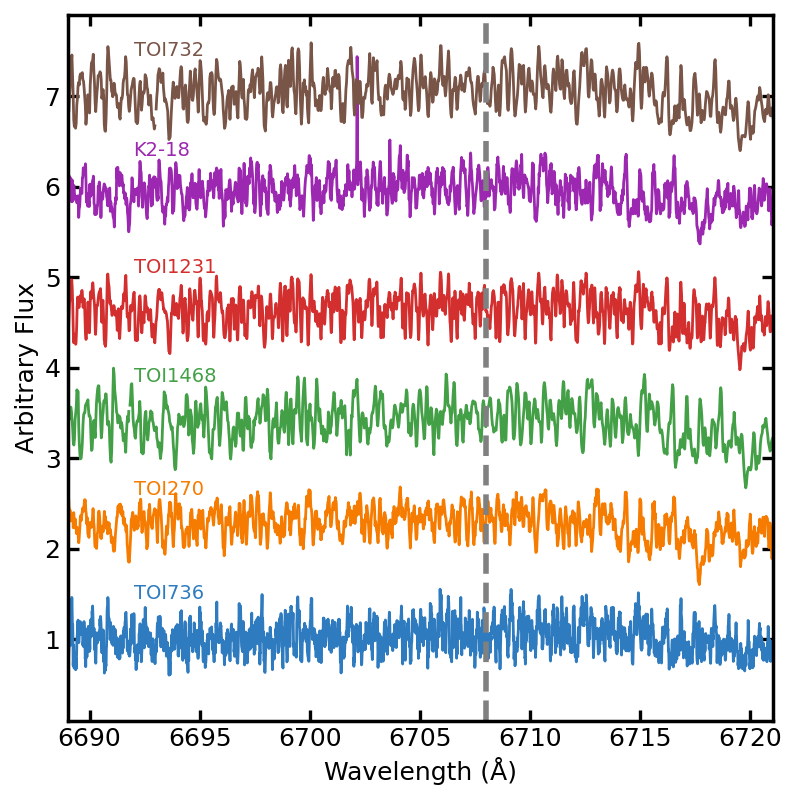}
\caption{Normalised spectra of our targets observed near the lithium absorption feature.  The data represent the highest signal-to-noise ratios for each target. The vertical dashed line at 6708 Å represents the location of the lithium absorption feature. }
\label{fig:lithium_abund}
\end{figure}

The equivalent width ($W_\lambda$) of the Li\,\textsc{i} doublet at 6707.76/6707.91\,\AA\ was measured by direct integration of the normalised flux over a $\pm0.3$\,\AA\ window centred on 6707.8\,\AA.
No significant absorption was detected in any system, and all measured equivalent widths are consistent with zero within their uncertainties.  
We therefore report \(3\sigma\) upper limits following the formalism of \citet{Cayrel1988},
\begin{equation}
\sigma_{\mathrm{EW}} \simeq 1.6\,\frac{\sqrt{\mathrm{FWHM}\,\delta x}}{\mathrm{SNR}},
\end{equation}
where \(\mathrm{FWHM} = \lambda / R\) represents the instrumental resolution element and \(\delta x\) denotes the pixel spacing near 6708\,\AA.  
The resulting measurements and upper limits for the highest-SNR exposure of each target are summarised in Table~\ref{tab:lithium_ew}.  
Figure~\ref{fig:lithium_abund} shows the corresponding spectra, with no detectable Li\,\textsc{i} absorption at 6708\,\AA\ (indicated by the vertical line), confirming substantial lithium depletion in all systems.

To interpret these limits, we compared our results with both theoretical and empirical lithium-depletion frameworks. Theoretical models of low-mass stellar evolution \citep[e.g.][]{Burrows_1997,Baraffe_2015} show that lithium is rapidly destroyed through proton capture once the stellar core becomes sufficiently hot, at temperatures of order $2.5\,\times 10^{6}$ K. In fully convective low-mass stars, efficient mixing between the core and the surface leads to depletion of surface lithium once this threshold is reached. For mid-to-late M dwarfs, lithium is expected to be exhausted within $\lesssim$100--300\,Myr, while higher-mass K and G dwarfs deplete lithium more gradually over longer timescales.

We further compared our measured equivalent widths and upper limits (Table~\ref{tab:lithium_ew}) with the empirical Li\,\textsc{i}\,6708\,\AA\ equivalent-width--temperature--age relations derived from the \textsc{EAGLES} model of \citet{Jeffries2023}. The \textsc{EAGLES} framework provides homogeneous lithium-depletion sequences for over 6000 stars in 52 open clusters, spanning ages from approximately 2\,Myr to 6\,Gyr and effective temperatures between 3000 and 6500\,K. For cool M dwarfs ($T_{\mathrm{eff}}\!\lesssim\!3800$\,K), both theoretical models and the empirical \textsc{EAGLES} sequences show that lithium absorption declines rapidly between the ages of young clusters such as the Pleiades ($\sim$125\,Myr) and intermediate-age clusters like Praesepe ($\sim$700\,Myr), becoming undetectable by a few hundred Myr. Our non-detections, all with $3\sigma$ limits of $W_\lambda \lesssim 30$\,m\AA, are therefore consistent with both theoretical predictions and empirical cluster sequences, indicating that all targets are older than the lithium-retention phase observed in young open clusters. Taken together, the empirical and theoretical comparisons indicate that all stars in our sample are older than the lithium-depletion boundary, implying lower age limits of approximately $\gtrsim$200\,Myr.

\begin{table*}
\centering
\caption{
Li\,\textsc{i}\,6708\,\AA\ equivalent-width measurements for the best-SNR spectrum of each target.
}
\begin{tabular}{lcccccccc}
\hline
Target & Instrument & $\delta x$ (Å) & SNR$_\mathrm{pix}$ & SNR$_\mathrm{res}$ & $W_\lambda$ (Å) & $\sigma_{\mathrm{EW}}$ (Å) & $3\sigma$ limit (mÅ)\\
\hline
TOI\,736  & ESPRESSO & 0.01 & 3.33 & 7.64 & 0.0000 & 0.0100 & 30.1 \\
TOI\,270  & HARPS    & 0.01 & 7.28 & 17.59 & 0.0080 & 0.0053 & 15.9 \\
TOI\,1468 & CARMENES & 0.03 & 7.17 & 11.47 & 0.0192 & 0.0099 & 29.7 \\
TOI\,1231 & ESPRESSO & 0.01 & 5.79 & 13.28 & 0.0078 & 0.0058 & 17.3 \\
K2--18    & HARPS    & 0.01 & 6.74 & 16.27 & 0.0033 & 0.0057 & 17.2 \\
TOI\,732  & CARMENES & 0.03 & 7.01 & 11.21 & 0.0053 & 0.0101 & 30.4 \\
\hline
\end{tabular}
\begin{flushleft}
\footnotesize
Notes.
SNR values are computed uniformly from the 6702–6706 and 6710–6714\,\AA\ continuum regions (Appendix~\ref{app:snr_calc}).  
$\delta x$ denotes the median pixel spacing near 6708\,\AA, and FWHM = $\lambda/R$.  
Equivalent widths were measured by direct integration over 6707.8\(\pm\)0.3\,\AA, and uncertainties estimated using the \citet{Cayrel1988}.
\end{flushleft}
\label{tab:lithium_ew}
\end{table*}

\subsection{Age constraints from rotation periods} \label{sec:rot_age}

The rotation--age relation adopted here follows \citet{Lalitha2025}, calibrated using open-cluster M dwarfs compiled from \citet{Curtis_2020} and \citet{Godoy-Rivera_2021}, with comparison to the field-star compilation of \citet{Engle2023}. The principal age anchors are the Pleiades (\(\sim\)0.12 Gyr), Praesepe (\(\sim\)0.67 Gyr), NGC~6811 (\(\sim\)1.0 Gyr), NGC~752 (\(\sim\)1.4 Gyr), Ruprecht~147 (\(\sim\)2.7 Gyr), and M67 (\(\sim\)4.0 Gyr). We note that spectral-type coverage is not uniform across age and is sparse for cooler M dwarfs at the oldest ages. We therefore treat rotation-based ages as empirical estimates conditioned on the available calibration set, with reduced robustness in sparsely sampled regions of the age--spectral-type plane.

We fit the relation as a third-order polynomial in
\(\log P_{\mathrm{rot}}\) using MCMC implemented with the emcee package. The adopted relation is
\begin{equation}
\log_{10}\left(\frac{\rm Age}{\rm Gyr}\right) = c_0 + c_1 x + c_2 x^2 + c_3 x^3,
\end{equation}

where \(x = \log_{10}(P_{\mathrm{rot}}/{\rm d})\) and \(c_0 - c_3\) are the polynomial coefficients. The relation is valid over \(P_{\mathrm{rot}} = 0.4\)--160 d. 
We also include an intrinsic-scatter term, \(\sigma_{\mathrm{int}}\) as an additional free parameter, representing the
star-to-star spread in excess of the quoted measurement
uncertainties. From the fit we obtain \(c_0 = -0.893\pm0.066\),
\(c_1 = 0.127\pm0.141\), \(c_2 = 0.768\pm0.220\),  \(c_3 = -0.201\pm0.072\) and \(\sigma_{\mathrm{int}} = 0.088^{+0.023}_{-0.020}\) dex.

All rotation-based ages quoted in this work, including those in
Table~\ref{tab:age_eg23}, are posterior credible intervals, obtained by
propagating a uniform prior over each adopted \(P_{\mathrm{rot}}\) range through
the relation and marginalising over both the polynomial coefficients and
\(\sigma_{\mathrm{int}}\). A uniform prior is used rather than merging the
individual datasets, for the reasons given in Appendix~\ref{app:rot_period}.

Figure~\ref{fig:age_rot_relation} shows the adopted rotation--age relation, obtained from our fit including an intrinsic-scatter term, which is more conservative than that obtained in \citet{Lalitha2025} without the intrinsic-scatter term.  The polynomial coefficients are
strongly correlated, so the uncertainty on the relation is better represented by predictive bands than by individual coefficient errors. Black markers show the \citet{Engle2023} sample, while cluster M dwarfs from \citet{Curtis_2020} and \citet{Godoy-Rivera_2021} are colour-coded by spectral type.

\begin{figure}
\includegraphics[width=0.5\textwidth]{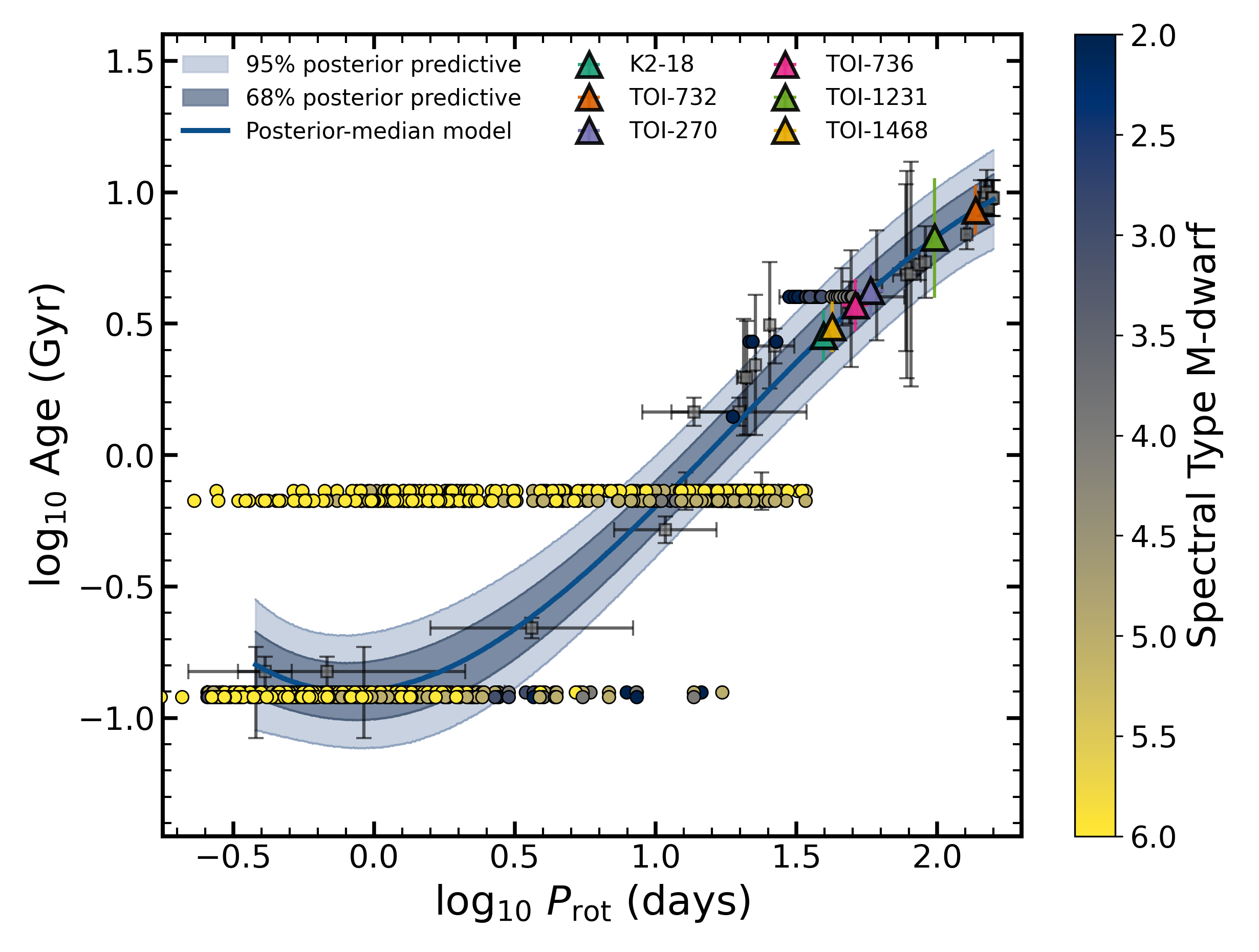}
\caption{Log--log relation between age and rotation period for M dwarfs. Black squares show the \citet{Engle2023} sample. Filled circles are cluster members from \citet{Curtis_2020} and \citet{Godoy-Rivera_2021}, colour-coded by spectral type (M2--M6). The solid blue curve is the median fit. Dark- and light-grey bands denote the 68\% and 95\% posterior predictive regions, including intrinsic astrophysical scatter. Triangles with error bars indicate the target stars analysed in this work.}
\label{fig:age_rot_relation}
\end{figure}

The calibrated rotation--age relation was then applied to the six M-dwarf planet hosts in our sample:

\begin{enumerate}
\item TOI-736: Using CATALINA and MEarth photometric data, we estimate a rotation period of 46--57 days (Appendix~\ref{app:rot_period}, Fig.~\ref{fig:rotation_toi736}). Applying the third-degree polynomial model gives an age range of $3.72^{+0.96}_{-0.76}$ Gyr.

\item TOI-270: \citet{Van_Eylen_2021} report a rotation period of 52--65 days from HARPS-based analysis. Propagating this interval through the adopted calibration gives an age range of $4.21^{+1.09}_{-0.86}$ Gyr.

\item TOI-1468: \citet{Chaturvedi2022} find a rotation period of 41--44 days from CARMENES and MAROON-X spectroscopy, supported by photometry. Our re-analysis of CARMENES and CATALINA data gives a consistent result (Appendix~\ref{app:rot_period}, Fig.~\ref{fig:rotation_toi1468}), corresponding to an age range of $3.07^{+0.77}_{-0.61}$ Gyr.

\item TOI-1231: A direct rotation period measurement is not available from existing time-series coverage. From ESPRESSO Ca\,\textsc{ii}\,H\&K spectra, we derive \(\log R'_{\mathrm{HK}}=-5.35\pm0.07\), corresponding to \(P_{\mathrm{rot}}=98.2\pm1.1\) days in the unsaturated regime (Appendix~\ref{app:rhk_prot}). Propagating this through the third-order rotation--age relation, with an intrinsic dispersion of \(\pm0.26\) dex in \(\log P_{\mathrm{rot}}\) estimated from local activity--rotation scatter (Fig.~\ref{fig:rhk_prot_plot}), yields \(6.70^{+4.57}_{-2.75}\,\mathrm{Gyr}\).

\item K2-18: \citet{Lalitha2025} report consistent photometric and spectroscopic estimates, with \(P_{\mathrm{rot}}=39\)--40 days. The corresponding age from the third-degree polynomial calibration is $2.84^{+0.70}_{-0.56}$ Gyr.

\item TOI-732: ASAS, CATALINA, and MEarth data give slightly different periods (143, 135, and 136 days). From the combined posterior distribution, \citet{Lalitha2025} infer \(P_{\mathrm{rot}}=130\)--145 days, giving an age range of $8.55^{+2.06}_{-1.64}$ Gyr.
\end{enumerate}

Overall, the rotation periods span 39--145 days, corresponding to ages of approximately 2.8--8.6 Gyr.

To check consistency with an alternative empirical parameterisation, we also derived ages using the early-M and late-M relations of \citet{Engle2023}, selecting the branch by spectral subtype. Table~\ref{tab:age_eg23} summarises the age intervals from both approaches.

\begin{table}
\centering
\caption{Rotation-based ages from this work compared with \citet{Engle2023}.
For each star we list the adopted rotation period range, the age derived from
our calibration, and the age interval obtained by mapping the same
$P_{\rm rot}$ range through the empirical relations of \citet{Engle2023}.}
\label{tab:age_eg23}
\begin{tabular}{lcccc}
\hline
Target & SpT & $P_{\rm rot}$  & Age$_\textrm{this work}$  & Age$_\textrm{E\&G 2023}$  \\
 &  & (days) & [Gyr] & [Gyr] \\
\hline
K2-18   & M2.5 & 39.0--40.0 & $2.84^{+0.70}_{-0.56}$ & 3.35--3.38 \\
TOI-732 & M4.0 & 130--145   & $8.55^{+2.06}_{-1.64}$ & 7.76--8.91 \\
TOI-270 & M3.0 & 52.0--65.0 & $4.21^{+1.09}_{-0.86}$ & 3.78--4.27 \\
TOI-736 & M6.0 & 46.0--57.0 & $3.72^{+0.96}_{-0.76}$ & 3.59--3.96 \\
TOI-1231& M3.0 & 97.1--99.3 & $6.70^{+4.57}_{-2.75}$ & 5.74--5.85 \\
TOI-1468& M3.5 & 41.0--44.0 & $3.07^{+0.77}_{-0.61}$ & 3.42--3.52 \\
\hline
\end{tabular}

\vspace{2mm}
\raggedright
Notes.
Ages from this work are posterior credible intervals (16th, 50th and 84th
percentiles). The \citet{Engle2023} column is a mapped range through their
relations and excludes intrinsic scatter. TOI-1231's period is inferred from $\log R'_{\rm HK}$ rather than measured, and carries the additional scatter of that relation.
\end{table}

\subsection{Kinematics constraint}

We began by calculating the Galactic space velocity components (U,V,
W) of our targets using its right ascension, declination, proper motions, distance, and radial velocity in Table\ref{tab:target}. These velocities were transformed into the Galactic coordinate system and corrected for the Solar motion relative to the Local Standard of Rest (LSR).

\begin{figure}
\includegraphics[width=0.5\textwidth]{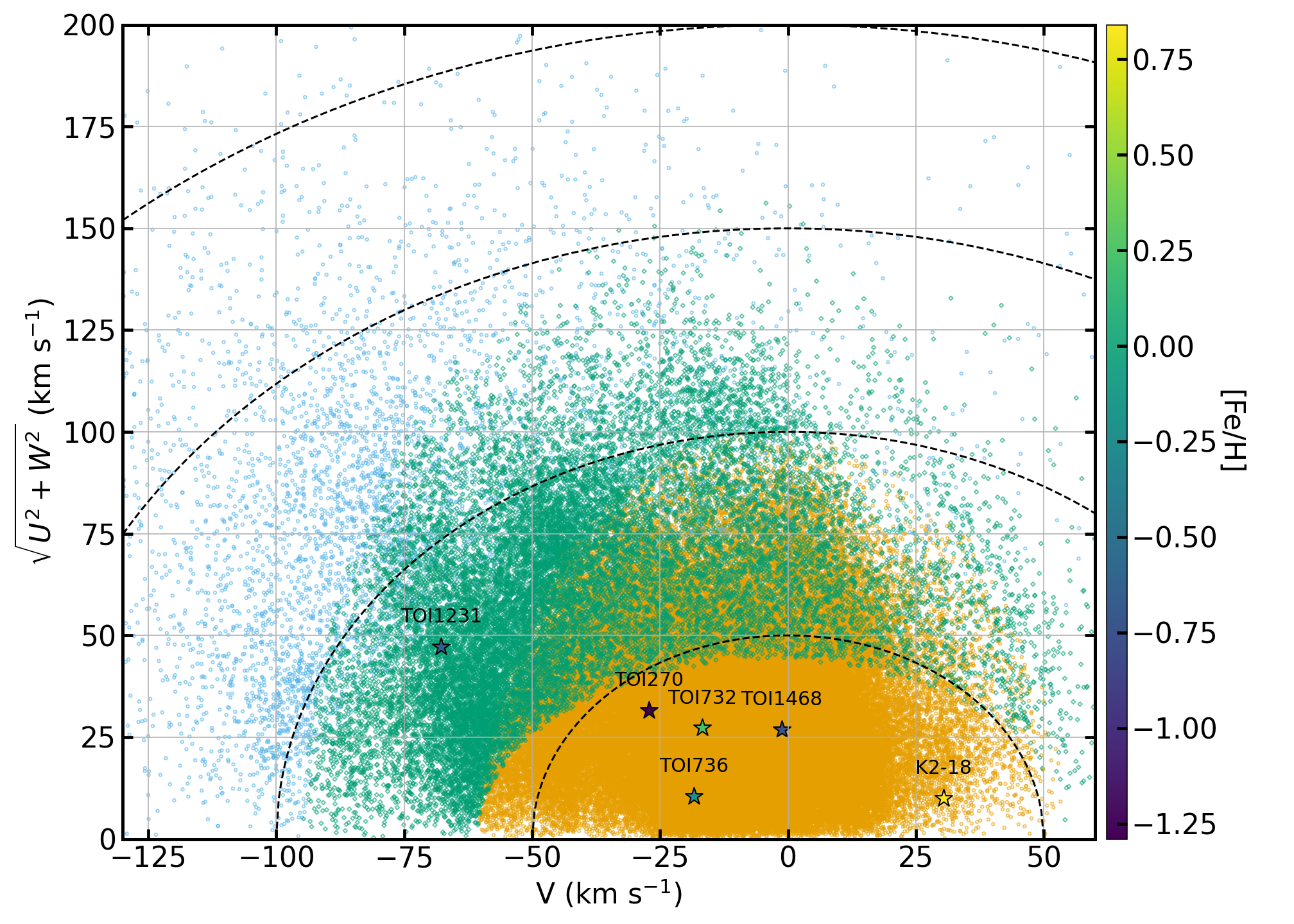}
\caption{Toomre diagram showing the total space velocity relative to the LSR, $\sqrt{U^2 + V^2 + W^2}$, plotted against the V velocity component for the target stars. Kinematic classification was further refined using the probabilistic method of \citet{Bensby2003}, with population assignments based on the ratio of thick-to-thin disk probabilities. The colour scale denotes stellar metallicity [Fe/H].}
\label{fig:Toomre}
\end{figure}

Galactic population membership is examined using a Toomre diagram. In Figure~\ref{fig:Toomre}, we plot the total velocity relative to the LSR, $\sqrt{U^2+V^2+W^2}$, against the rotational velocity component \(V\). The diagram enables a statistical distinction between the Galactic thin disk, thick disk, and halo populations. Stars within the thin disk exhibit low total velocities and rotational velocities close to the LSR, while thick disk and halo stars show progressively higher total velocities and distinct kinematic patterns.

In addition to the Toomre diagram, we utilized the kinematic probability method of \cite{Bensby2003} to quantitatively classify each star's Galactic membership. This approach uses Gaussian velocity distributions for the thin disk, thick disk, and halo populations, incorporating their respective velocity dispersions ($\sigma_U$, $\sigma_V$, $\sigma_W$), asymmetric drift, and relative population fractions. Probabilities for each population were calculated and normalised, yielding the fractional likelihood of a star's membership in the thin disk (P$_{\emph{thin}}$), thick disk (P$_{\emph{thick}}$), or halo (P$_{\emph{halo}}$).

The ratio of thick disk to thin disk probabilities (P$_{\emph{thick}}$/P$_{\emph{thin}}$) served as the primary criterion for classification:
\begin{itemize}
    \item Stars with P$_{\emph{thick}}$/P$_{\emph{thin}}$$\geq$ 10 were classified as thick disk members.
    \item Stars with P$_{\emph{thick}}$/P$_{\emph{thin}}$$\leq$ 0.1 were classified as thin disk members.
\end{itemize}

We further incorporated stellar metallicities ([Fe/H]) to refine the kinematic classifications. Each target was colour-coded by [Fe/H] in the Toomre diagram (Figure~\ref{fig:Toomre}), allowing a direct comparison between chemical composition and Galactic dynamics. For borderline cases (\(0.1 < P_{\mathrm{thick}}/P_{\mathrm{thin}} < 10\)), metallicity provided an additional discriminant: metal-rich stars ([Fe/H]~$\gtrsim$~--0.3) were assigned to the thin disk, while metal-poor stars ([Fe/H]~$\lesssim$~--0.5) were considered thick-disk--like.  

This combined kinematic and metallicity analysis confirms that all targets are consistent with thin-disk membership, with the exception of TOI-1231, which lies near the thin/thick-disk boundary (Table~\ref{tab:coefficients}).

\begin{table}
    \centering
    \caption{Kinematic association percentages for thin-disk, thick-disk, and halo populations for our targets.}
    \begin{tabular}{ccccc}
        \hline
        Target & P$_{\mathrm{thin}}$ (\%) & P$_{\mathrm{thick}}$ (\%)  & P$_{\mathrm{halo}}$ (\%)  & Classification \\
        \hline
        TOI-736  & 99.35 & 0.64  & 1E-5 & Thin \\
        TOI-270  & 98.86 & 1.13  & 2E-4 & Thin \\
        TOI-1468 & 98.99 & 1.00  & 2E-4 & Thin \\
        TOI-1231 & 13.14 & 86.76 & 9E-2 & Uncertain \\
        K2-18    & 99.31 & 0.69  & 1E-5 & Thin \\
        TOI-732  & 98.06 & 1.93  & 4E-4 & Thin \\
        \hline
    \end{tabular}

Notes: Values are percentages. Uncertain denotes borderline thin/thick-disk classification.
\label{tab:coefficients}
\end{table}

\begin{table*}
    \centering
    \setlength{\tabcolsep}{3.0pt}
    \caption{Kinematical ages obtained through the methods UVW (Method 1), eVW (Method 2), and eUW (Method 3) following \citealt{Almeida2018}}
    \label{tab:age_kinematics}
    \begin{tabular}{l|llllll|llllll|llllll}
    \hline
    Star & \multicolumn{6}{|c|}{Method 1 (UVW)} & \multicolumn{6}{c|}{Method 2 (eVW)} & \multicolumn{6}{c}{Method 3 (eUW)} \\
    & t$_{\mathrm{E}}$ & t$_{\mathrm{ML}}$ & t$_{16\%}$ & t$_{50\%}$ & t$_{84\%}$ & t$_{\mathrm{Kin}}$ 
    & t$_{\mathrm{E}}$ & t$_{\mathrm{ML}}$ & t$_{16\%}$ & t$_{50\%}$ & t$_{84\%}$ & t$_{\mathrm{Kin}}$ 
    & t$_{\mathrm{E}}$ & t$_{\mathrm{ML}}$ & t$_{16\%}$ & t$_{50\%}$ & t$_{84\%}$ & t$_{\mathrm{Kin}}$ \\
    & Gyr & Gyr & Gyr & Gyr & Gyr & Gyr & Gyr & Gyr & Gyr & Gyr & Gyr & Gyr & Gyr & Gyr & Gyr & Gyr & Gyr&Gyr \\
    \hline
    TOI-736 & 2.82  & 0.10  & 0.29  & 1.41  & 6.04  & $0.78^{+5.26}_{-0.48}$  
            & 4.12  & 1.21  & 1.16  & 2.99  & 7.68  & $1.94^{+5.74}_{-0.78}$  
            & 4.82  & 1.10  & 1.34  & 3.79 & 9.02  & $2.03^{+6.99}_{-0.69}$  \\

    TOI-270 & 6.21  & 2.73  & 2.42  & 5.68  & 10.37  & $3.60^{+6.77}_{-1.18}$  
            & 6.49  & 2.99  & 2.69  & 5.97  & 10.72  & $3.87^{+6.85}_{-1.18}$  
            & 6.16  & 2.08  & 2.30  & 5.54  & 10.52  & $3.10^{+7.43}_{-0.80}$  \\

    TOI-1468 & 5.30  & 2.10  & 1.98  & 4.55 & 9.03  & $2.90^{+6.13}_{-0.92}$  
             & 7.60  & 5.37  & 3.90  & 7.39 & 11.55  & $5.93^{+5.62}_{-2.03}$  
             & 5.93  & 2.42  & 2.38  & 5.19 & 10.02  & $3.30^{+6.72}_{-0.92}$  \\

    TOI-1231 & 11.02  & 14.00  & 8.66  & 11.45 & 13.28  & $13.26^{+0.02}_{-4.60}$  
             & 10.80  & 13.69  & 8.30  & 11.20 & 13.19  & $12.97^{+0.22}_{-4.68}$  
             & 10.90  & 13.92  & 8.46  & 11.34 & 13.23  & $13.16^{+0.07}_{-4.70}$  \\

    K2-18    & 3.90  & 0.46  & 0.74  & 2.72 & 7.75  & $1.32^{+6.43}_{-0.58}$  
             & 5.15  & 1.98  & 1.78  & 4.22 & 9.13  & $2.77^{+6.36}_{-0.99}$  
             & 5.20  & 1.55  & 1.62  & 2.36 & 9.46  & $2.46^{+7.00}_{-0.84}$  \\

    TOI-732  & 6.85  & 4.23  & 3.23  & 6.49 & 10.74  & $4.89^{+5.86}_{-1.66}$  
             & 8.23  & 6.83  & 4.71  & 8.16 & 11.91  & $7.18^{+4.73}_{-2.48}$  
             & 8.04  & 6.61  & 4.40  & 7.95 & 11.86  & $6.97^{+4.89}_{-2.57}$  \\

    \hline
    \end{tabular}
    
    \footnotesize{Note: t$_{\mathrm{E}}$ is the expected age in Gyr; t$_{\mathrm{ML}}$ is the most likely age in Gyr;  
    t$_{16\%}$, t$_{50\%}$, and t$_{84\%}$ are the 16th, 50th (median), and 84th percentile ages in Gyr;  
    t$_{\mathrm{Kin}}$ is the kinematical age as weighted by average between t$_{\mathrm{ML}}$ and t$_{\mathrm{E}}$.}

\end{table*}

\subsection{Age Constraint from Kinematics}

In this study, we utilise three distinct methods to estimate stellar ages based on kinematic data: Method 1 (Bayesian Age Distribution), Method 2 (Eccentricity-Based Estimation using \( V \) and \( W \) velocities), and Method 3 (Eccentricity-Based Estimation using \( U \) and \( W \) velocities). These methods build upon the framework introduced by \citet{Almeida2018}, who derived stellar ages from space motions by modelling how the velocity dispersions of stellar populations (\(\sigma_U\), \(\sigma_V\), \(\sigma_W\)) evolve with age. For an individual star, the measured space velocities \(U\), \(V\), and \(W\) are compared against these age-dependent population dispersions to infer a probabilistic age.

\subsubsection{Method 1: Bayesian Age Distribution}

The first method calculates the age probability distribution using the star’s velocity components \( U \), \( V \), and \( W \), along with the star's kinematic parameters. These kinematic parameters are age-dependent, and we model the velocity dispersions in each of the three directions (\( U \), \( V \), and \( W \)) as a function of age.

For the velocity dispersions, we use the age-dependent relations from \citet{Almeida2018}, where $\sigma_U$, $\sigma_V$, and $\sigma_W$ are in km,s$^{-1}$ and age is expressed in Gyr:

\begin{equation}
\sigma_U(\text{age}) = 21.2\pm1.0 \times (\text{age})^{0.35\pm0.02}
\end{equation}
\begin{equation}
\sigma_V(\text{age}) = 13\pm1.0 \times (\text{age})^{0.36\pm0.02}
\end{equation}
\begin{equation}
\sigma_W(\text{age}) = 9.1\pm1.0 \times (\text{age})^{0.48\pm0.04}
\end{equation}

These relations were calibrated by \citet{Almeida2018} using Gaia DR2 kinematics of nearby FGKM stars. The same relations were applied to the M8 dwarf TRAPPIST-1, yielding a kinematic age consistent with the independent estimate of \citet{Burgasser2017}, indicating that they remain applicable for low-mass stars such as those considered in this work.

The stellar space motion relative to the Sun is also taken into account, using a transformation for the velocities \( U \), \( V \), and \( W \) as follows:

\begin{equation}\label{eqn:v_1}
v_1 = (U + U_{\odot}) \cos(l_v) + (V + V_{\odot}) \sin(l_v) 
\end{equation}
\begin{equation}\label{eqn:v_2}
v_2 = -(U + U_{\odot}) \sin(l_v) + (V + V_{\odot}) \cos(l_v)
\end{equation}
\begin{equation}\label{eqn:v_3}
v_3 = W + W_{\odot} 
\end{equation}

where \( l_v \) is the velocity angle, and \( U_{\odot} \), \( V_{\odot} \), and \( W_{\odot} \) represent the motion of the Sun. The likelihood function is based on these transformed velocities and their corresponding dispersions. The posterior distribution for age is given by:

\begin{equation}
p(t|U,V,W) \propto p(v_1|t) p(v_2|t) p(v_3|t) p(t)
\end{equation}

The corrections depend on age, particularly in the V-direction due to asymmetric drift. The velocity angle and age-dependent solar motion are modelled as follows:

\begin{equation}
l_v(t) = 0.41 \exp(-0.37 t)
\end{equation}
\begin{equation}
V_{\odot}'(t) = 0.17 t^2 + 0.63 t + 12.5
\end{equation}

The individual velocity components' likelihood is given by:

\begin{equation}
p(t|U,V,W) \propto \prod_{i=1,2,3} \frac{1}{\sqrt{2 \pi \sigma_i^2(t)}} \exp\left(-\frac{v_i^2}{2 \sigma_i^2(t)}\right)
\end{equation}

where \(v_1\), \(v_2\), and \(v_3\) are obtained from Equations~\ref{eqn:v_1}–\ref{eqn:v_3}, and \(\sigma_1(t)\), \(\sigma_2(t)\), and \(\sigma_3(t)\) correspond to the age-dependent velocity dispersions \(\sigma_U(t)\), \(\sigma_V(t)\), and \(\sigma_W(t)\), respectively.

\subsubsection{Method 2: Eccentricity-Based Estimation (eVW Method)}

In this method, we estimate the stellar age by incorporating orbital eccentricity along with the velocity components \(V\) and \(W\). The probability is given by:

\begin{equation}
p(t|e,V,W) = p(e|V,t) p(V|t) p(W|t) p(t)
\end{equation}

The eccentricity \(e\) is computed as:

\begin{multline}
   e = 2.98 \times 10^{-3} \\\sqrt{155 + 20.0 \times U + 19.6 \times V + U^2 + 1.95 \times V^2} \\- 7.23 \times 10^{-4}
\end{multline}

The velocity probability distribution is:

\begin{equation}
p(V|t) \cdot p(W|t) = \frac{1}{2\pi} \prod_{K=V,W} \left[ \frac{1}{\sigma_k(t)} \exp\left( -\frac{[k + k_{\odot}(t)]^2}{2 \sigma_k^2(t)} \right) \right]
\end{equation}

\subsubsection{Method 3: Eccentricity-Based Estimation (UW Method)}

In this method, we estimate the stellar age by incorporating the eccentricity, using the \(U\) and \(W\) velocity components. The probability is given by:

\begin{equation}
p(t|e,U,W) = p(e|U,t) p(U|t) p(W|t) p(t)
\end{equation}

The velocity probability distribution is:

\begin{equation}
p(U|t) \cdot p(W|t) = \frac{1}{2\pi} \prod_{K=U,W} \left[ \frac{1}{\sigma_k(t)} \exp\left( -\frac{[k + k_{\odot}(t)]^2}{2 \sigma_k^2(t)} \right) \right]
\end{equation}

The posterior distribution is normalised over the age range, and the most likely age and its uncertainties are derived.

For all methods, we compute three age estimates: the most likely age, \( t_{\text{ML}} \), defined as the mode of the posterior distribution; the expected age, \( t_{\text{E}} \), given by the posterior mean; and the kinematic age, \( t_{\text{kin}} \), derived from the velocity distribution following  \citet{Almeida2018}. We use the cumulative distribution of the posterior to estimate the 16th, 50th, and 84th percentiles of the age distribution, which provide the uncertainty bounds for each age estimate:

\begin{equation}
t_{\text{kin}} = \frac{3 t_{\text{ML}} + t_{\text{E}}}{4}
\end{equation}

The uncertainty is calculated by finding the difference between the kinematic age and the 16th and 84th percentiles, which gives the lower and upper uncertainty bounds. The resulting age estimates from all three methods, along with their associated uncertainties, are listed in Table~\ref{tab:age_kinematics}.

\begin{figure*}
\includegraphics[width=0.33\textwidth]{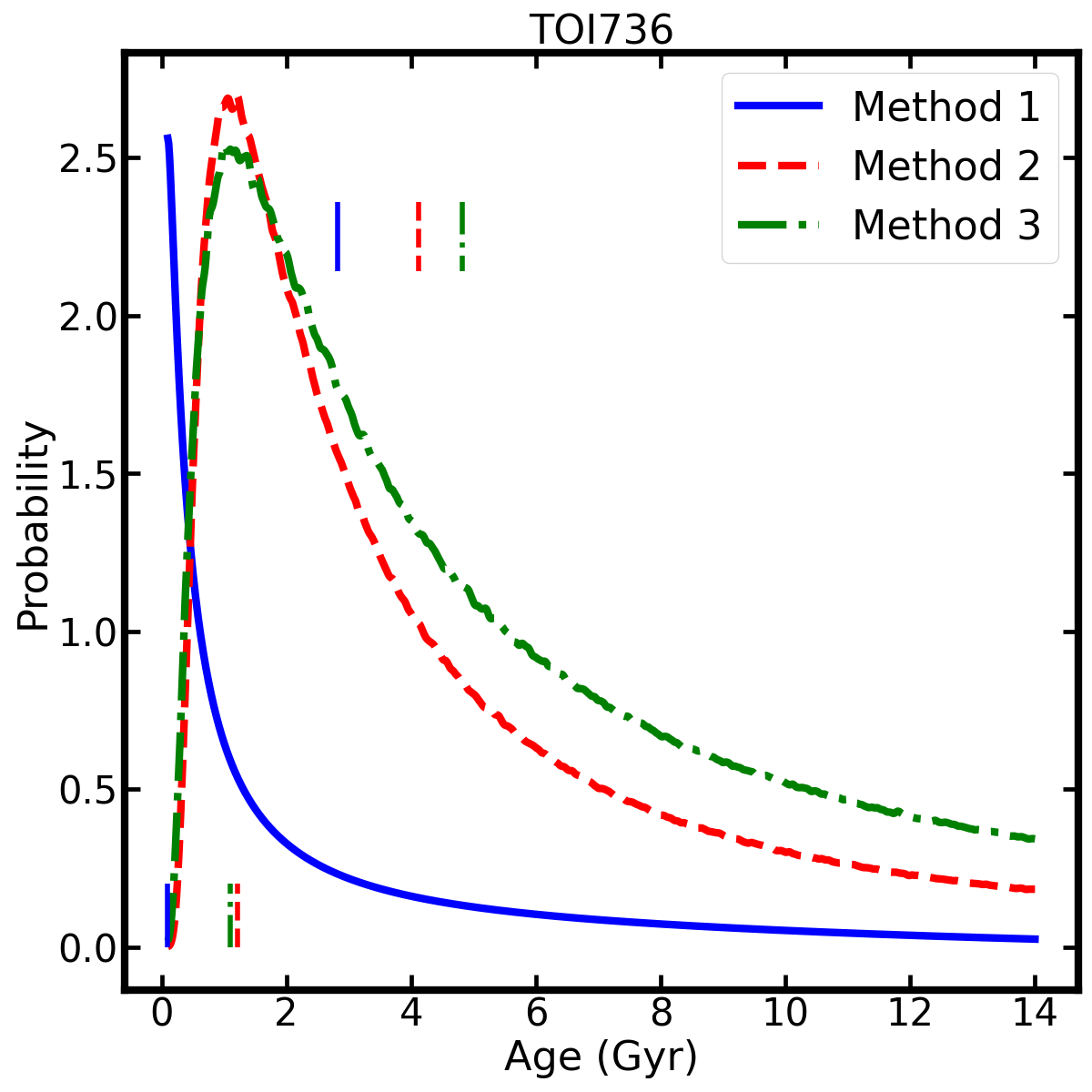}
\includegraphics[width=0.33\textwidth]{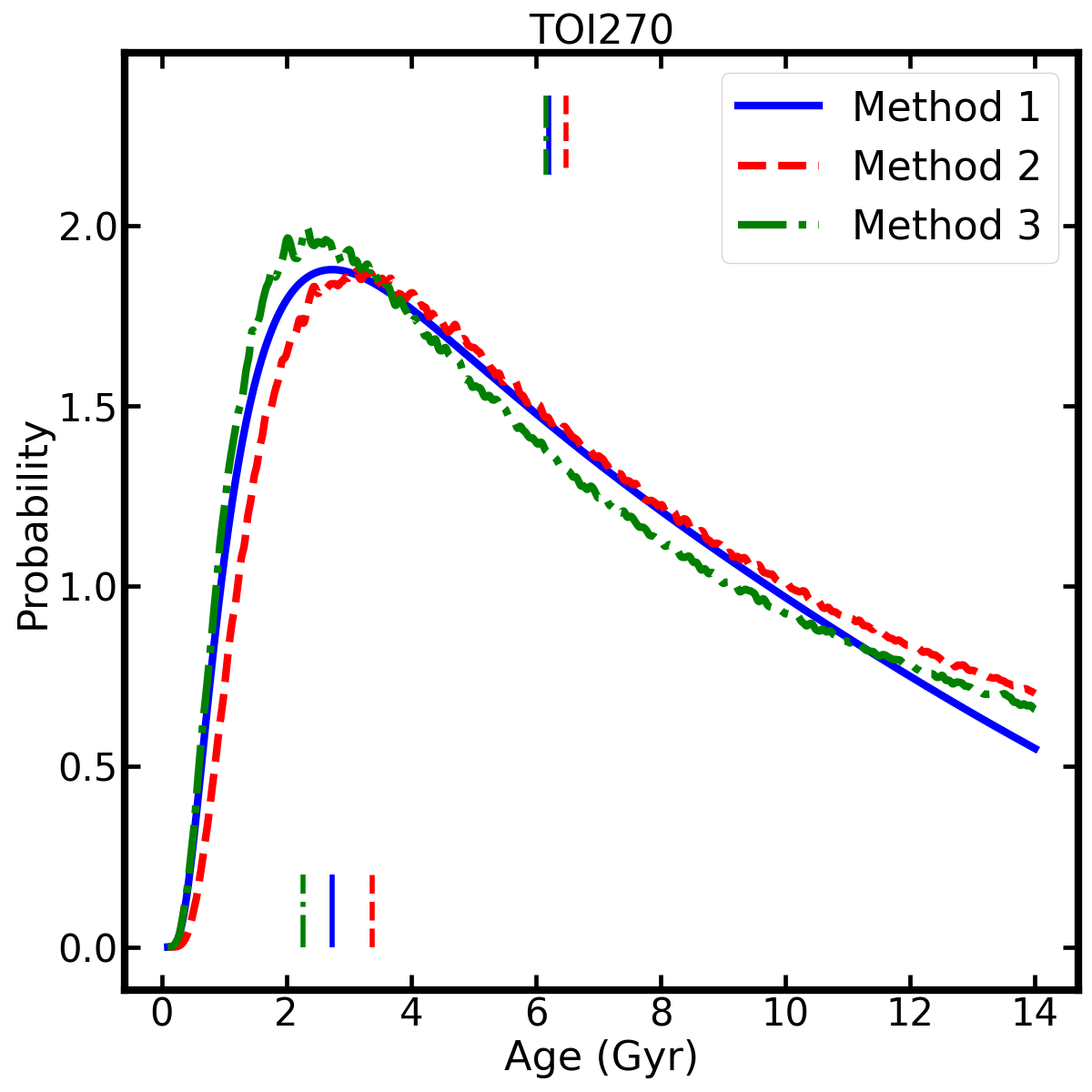}
\includegraphics[width=0.33\textwidth]{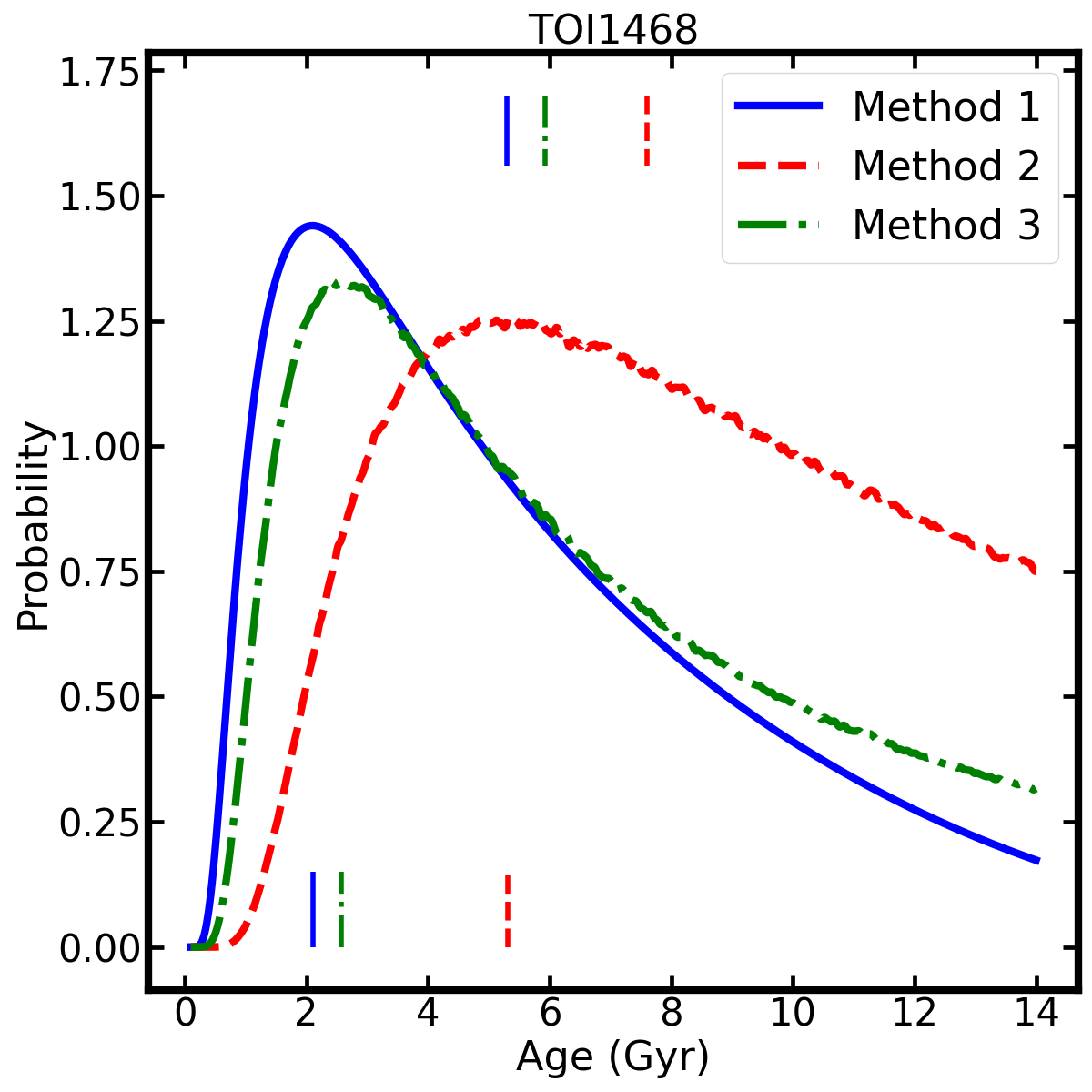}
\includegraphics[width=0.33\textwidth]{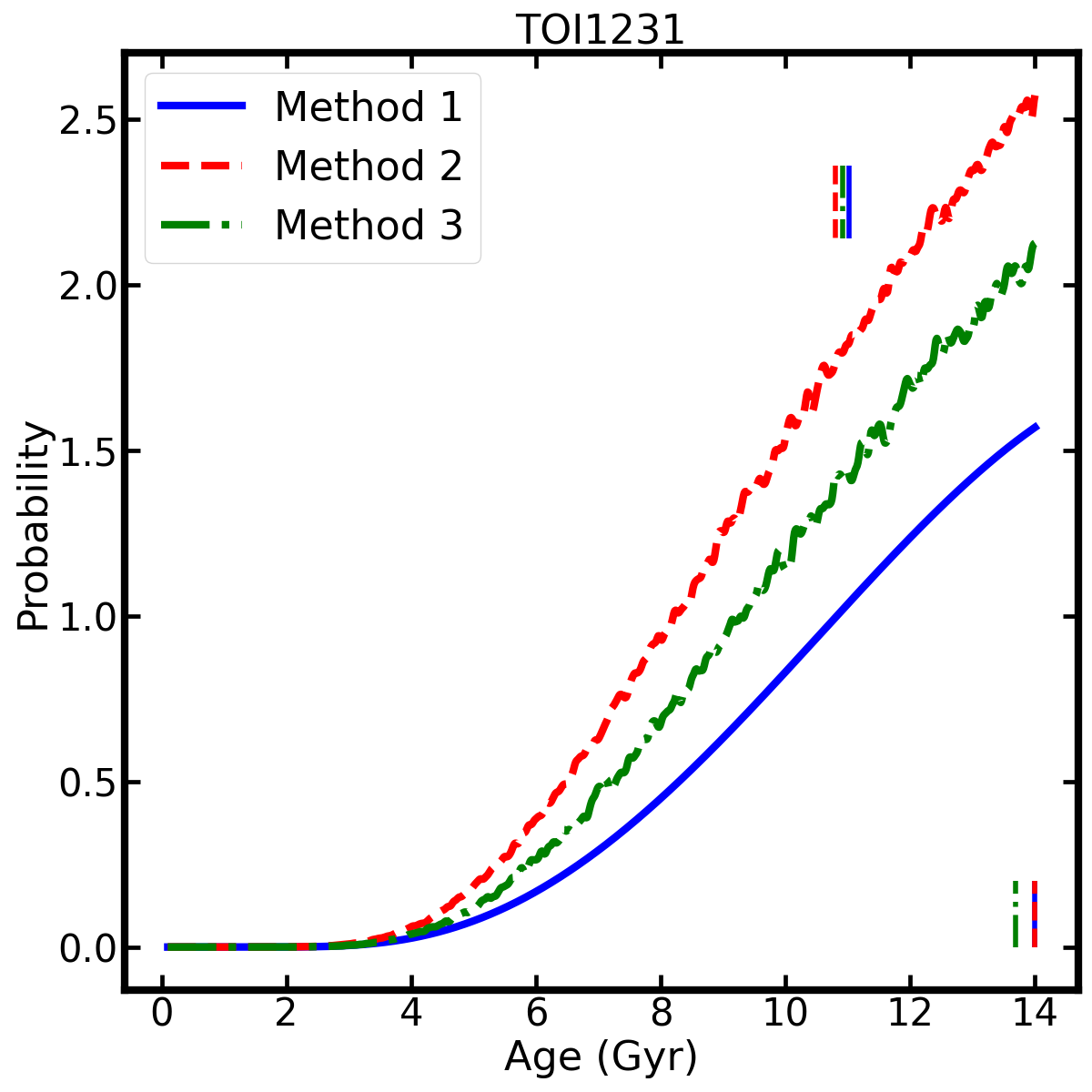}
\includegraphics[width=0.33\textwidth]{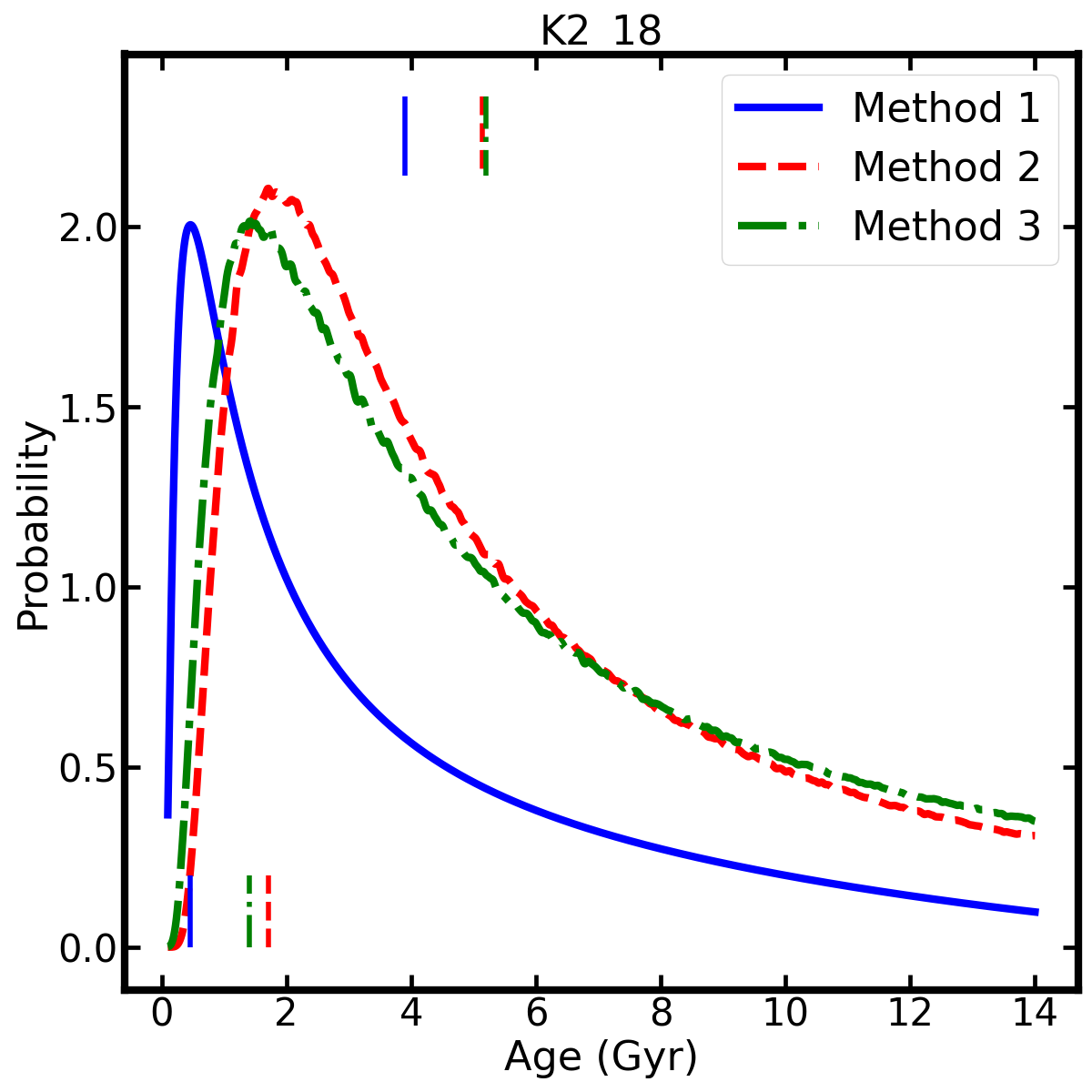}
\includegraphics[width=0.33\textwidth]{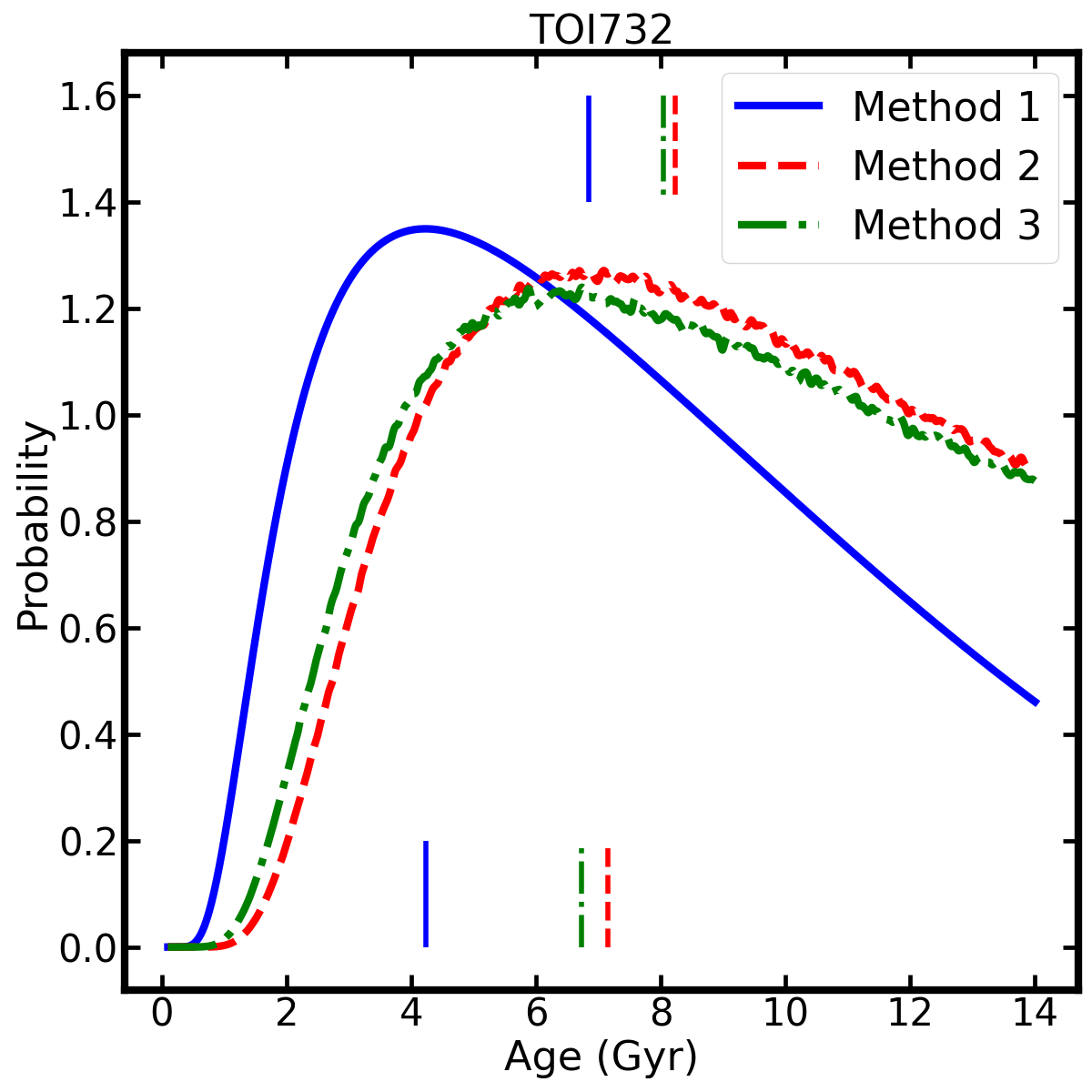}
\caption{Probability density distributions of the kinematic ages for six stars. Each plot displays the age likelihoods derived from different methods (Method 1 as a solid line, Method 2 as a dashed line, and Method 3 as a dotted-dashed line) for each star. The vertical ticks on the bottom axis represent the most likely age for each method, while those on the top axis correspond to the expected ages.}
\label{fig:kinmatic}
\end{figure*}

In this study, we employ three distinct methods for estimating the ages of our target stars, using their kinematic properties. The kinematic age estimates derived from these methods are displayed in Figure~\ref{fig:kinmatic}, where the probability density distributions of the kinematic ages are plotted for each star. As shown in Figure~\ref{fig:kinmatic}, the age likelihood distributions for six stars are presented, with Method 1 (solid line) typically predicting younger ages, while Method 2 (dashed line) and Method 3 (dotted-dashed line) predict older ages. The vertical ticks on the top of each plot represent the most likely age for each method, and the ticks at the bottom represent the expected age. 

To validate our methods, we also compared our age estimates to previously reported results for the Trappist-1 system. According to \cite{Almeida2018}, the most likely age for Trappist-1 
is 12.50$^{+0.29}_{-6.23}$ Gyr. Using our method, we obtained a most likely age of 12.88$^{+0.02}_{-6.27}$ Gyr. While our most likely age is slightly higher, the uncertainty ranges are very similar, with both methods providing consistent results within the uncertainties. Additionally, our expected age was found to be 12.50 Gyr, which is in agreement with Almeida et al.'s expected age for Trappist-1. This comparison confirms the validity of our implementation and shows that the results are in good agreement with previous studies.

\section{ Results and Discussion} \label{sec:results_discussion}

\begin{figure*}
\includegraphics[width=\textwidth]{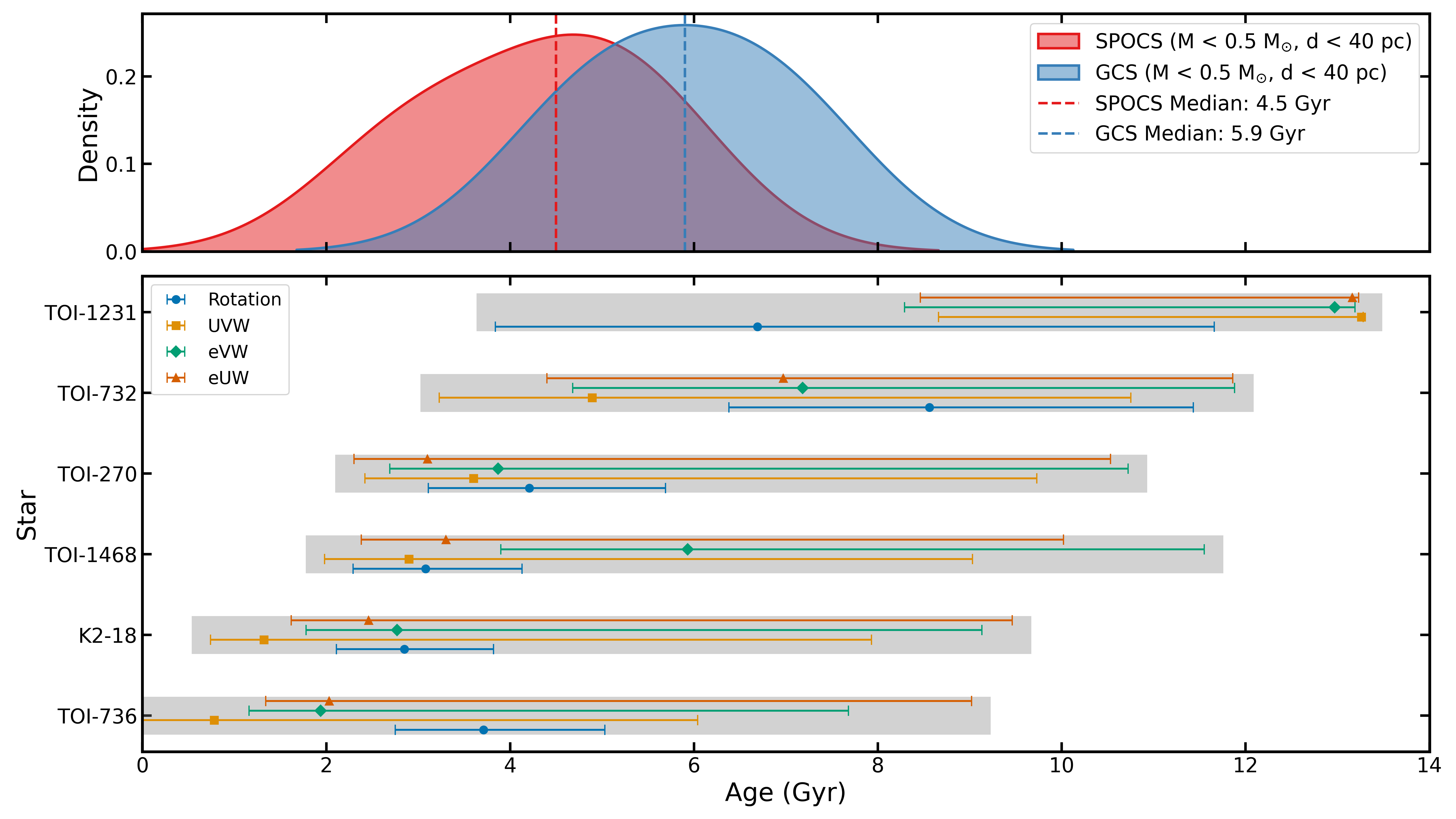}
\caption{Top panel: Age distribution of low-mass stars ($M < 0.5\,M_\odot$, $d < 40$ pc) from the SPOCS (red) and GCS (blue) samples, with median ages marked by dashed lines.
Bottom panel: Age estimates for the six sub-Neptune host stars from four methods: rotation (Table~\ref{tab:age_eg23}) and the kinematic UVW, eVW and eUW methods (Table~\ref{tab:age_kinematics}). Coloured points mark the median age from each method and the horizontal error bars their 16th–84th percentile credible intervals. The rotation-based intervals include the intrinsic dispersion of the rotation–age relation and are therefore directly comparable with the kinematic intervals. The grey region for each target spans the full range covered by all four methods.}
\label{fig:summary_plot}
\end{figure*}

We compare rotation-based and kinematic age estimates to assess the evolutionary status of the target stars. Figure~\ref{fig:summary_plot} summarises the results for our sample. The top panel shows the age distribution of nearby low-mass stars from the SPOCS (\textit{Spectroscopic Properties of Cool Stars}; \citealt{Valenti_2005}) and GCS (\textit{Geneva–Copenhagen Survey}; \citealt{Casagrande2011}) catalogues. The SPOCS catalogue provides spectroscopically derived stellar parameters and isochrone-based ages for nearby FGKM dwarfs, while the GCS catalogue includes kinematic and metallicity measurements for a large sample of solar neighbourhood stars, with ages determined from photometric isochrone analysis. Together, these catalogues provide a representative benchmark for the local stellar population, allowing a direct comparison between our M-dwarf age estimates and the broader distribution of field-star ages. The bottom panel presents age estimates for each planet-hosting star using different methods (rotation periods, UVW space velocities, and orbital eccentricities). For each star, the grey horizontal bar illustrates the ensemble of age estimates obtained from all diagnostic methods, while the coloured symbols with error bars denote the individual estimates derived from each method. Including the intrinsic dispersion of the rotation--age relation, the rotation-based ages have uncertainties of approximately 0.6--1.9 Gyr, compared with 2.3--3.8 Gyr for the kinematic estimates. For the five stars with directly measured rotation periods the rotation-based constraint is tighter by factors of 2--6. TOI-1231 is the exception, as its period is inferred from $\log R'_{\rm HK}$ rather than measured, and the scatter of that relation makes its rotation-based age comparable to its kinematic constraint.

\begin{itemize}

\item TOI-736: Using \(P_{\mathrm{rot}}=46\)--57 d, the rotation age is $3.72^{+0.96}_{-0.76}$~Gyr. The kinematic methods are much broader and favour an intermediate range of \(\sim\)1--4 Gyr. Owing to the large kinematic uncertainties, older ages (including \(\sim\)7--10 Gyr) remain possible in the low-probability tail, but are not preferred by either the rotation-based estimate or the peak kinematic probability.

\item TOI-270: For \(P_{\mathrm{rot}}=52\)--65 d, the  rotation age is $4.21^{+1.09}_{-0.86}$~Gyr. Kinematic estimates are broader, with most probable values around \(\sim\)2--4 Gyr and extended probability toward older ages. We therefore classify TOI-270 as most likely middle-aged, while noting the wider kinematic allowance.

\item TOI-1468:
For \(P_{\mathrm{rot}}=41\)--44 d, the  rotation age is $3.07^{+0.77}_{-0.61}$~Gyr. Kinematic methods 1 and 3 favour \(\sim\)2--4 Gyr, whereas method 2 tends towards older age; the combined kinematic range is therefore wider than the rotation range. TOI-1468 is most consistent with an intermediate age.

\item TOI-1231:
Using the activity-derived rotation period and full propagation of local intrinsic scatter, we obtain a rotation-based age of $6.70^{+4.57}_{-2.75}$~Gyr. All three kinematic methods place TOI-1231 at the older end of the sample, with characteristic kinematic ages near \(\sim\)13 Gyr. Taken together, TOI-1231 is the oldest system in our sample.

\item K2-18:
For \(P_{\mathrm{rot}}=39\)--40 d, the rotation age is $2.84^{+0.70}_{-0.56}$~Gyr. The kinematic posterior is broad but weighted toward younger ages than most of the other targets. K2-18 is therefore among the younger systems in this sample.

\item TOI-732:
For \(P_{\mathrm{rot}}=130\)--145 d, the rotation age is $8.55^{+2.06}_{-1.64}$~Gyr. Kinematic methods also favour moderate-to-old ages, although with broad uncertainty. TOI-732 is thus consistently identified as an old system by both diagnostics.

\end{itemize}

\section{Summary and conclusion} \label{sec:summary}

In this study, we present a comprehensive age analysis of six M dwarf exoplanet host stars: TOI-736, TOI-270, TOI-1468, TOI-1231, K2-18, and TOI-732. These systems host temperate sub-Neptune planets and are of particular interest for current and upcoming JWST observations aimed at atmospheric characterisation. However, determining the ages of M dwarfs remains a fundamental challenge, owing to their slow structural evolution and extended magnetic activity. Stellar age is a critical parameter for placing these planets in their evolutionary context, as it informs models of atmospheric escape, photochemistry, and interior processes, and is essential for interpreting atmospheric data from JWST and complementary observational efforts.

We combined multiple independent age diagnostics, including lithium absorption, stellar rotation periods, and Galactic kinematics, to constrain the ages of the host stars. Our main findings are as follows:
\begin{itemize}
    \item No detectable Li~I absorption was observed in the 6708~\AA\ region for any target. This non-detection indicates significant lithium depletion, consistent with stellar ages exceeding $\sim$200 Myr based on theoretical models.
    \item Rotation periods, derived from photometric and spectroscopic observations, range from approximately 39 to 145 days. Using a calibrated third-degree polynomial model, we estimated gyrochronological ages between 2.8 and 8.6 Gyr.
    \item Kinematic age estimates, obtained using three methods based on velocity dispersion and orbital eccentricity, provided consistent but broad age ranges. These methods confirmed that TOI-1231 is the oldest system in our sample, with a likely age close to 13 Gyr. The kinematic estimates favour the youngest ages for TOI-736 and K2-18, at 0.8 and 1.3 Gyr respectively, although with uncertainties of several Gyr these are not preferred over the older rotation-based ages.
\end{itemize}

However, the significant scatter between methods, especially in the kinematic estimates, highlights the inherent uncertainties and the limitations of relying on any single diagnostic. While rotation periods tend to align with older kinematic ages in some cases, discrepancies remain, particularly for stars with dynamically cold or eccentric orbits. These results demonstrate the importance of combining multiple independent indicators and careful interpretation of age diagnostics in M dwarfs.

Among the systems analysed here:
\begin{itemize}
    
    \item K2-18 shows relatively short rotation periods and low space velocities, consistent with thin disk membership and a relatively young age.
    \item TOI-736, TOI-270, and TOI-1468 exhibit intermediate-age characteristics, with rotation periods and kinematic indicators suggesting ages between $3-5$ Gyr.
    \item TOI-1231 and TOI-732 appear to be the most evolved, exhibiting long rotation periods and kinematics consistent with older stellar populations. TOI-1231, in particular, shows strong evidence for membership in the thick disk and represents one of the oldest stars in our sample.
    
\end{itemize}

Our results show that rotation provides the tighter age constraint for stars with directly measured rotation periods, by factors of 2--6, while for stars whose period must be inferred from chromospheric activity the two approaches are comparable. They remain complementary, with kinematics providing independent checks, identifying extreme-age cases, and placing the stars in their Galactic context. The combination yields a more complete view of the evolutionary status of M dwarf exoplanet hosts, despite the broad uncertainties inherent to kinematic methods.

Methodological improvements will further strengthen these diagnostics. Long-term rotation monitoring and refined gyrochronology calibrations will improve rotation-based ages, while updated velocity–age relations from surveys such as \textit{Gaia} will reduce systematic uncertainties in kinematic estimates. Together, these advances will enable more reliable and informative ages for low-mass stars.

Thus, combining rotation and kinematic diagnostics provides a consistent approach to age determination in M dwarf exoplanet hosts and establishes a basis for future refinement with larger samples.

\section*{Acknowledgements}

This work is supported by the UK Research and Innovation (UKRI) Frontier Research Grant EP/X025179/1 (PI: Nikku Madhusudhan). We thank Ted von Hippel for discussion and for providing feedback on the manuscript.

\section*{Data Availability}

The spectroscopic data analysed in this study are publicly available from the following archives: 
HARPS and ESPRESSO can be downloaded from the ESO Science Archive\footnote{\href{http://archive.eso.org/}{http://archive.eso.org/}}; 
CARMENES data are accessible from the CARMENES archive\footnote{\href{http://carmenes.cab.inta-csic.es/gto/jsp/dr1Public.jsp}{http://carmenes.cab.inta-csic.es/gto/jsp/dr1Public.jsp}}. 
Photometric data are available from the ASAS\footnote{\href{https://www.astrouw.edu.pl/asas/}{https://www.astrouw.edu.pl/asas/}}, 
CATALINA\footnote{\href{http://nesssi.cacr.caltech.edu/DataRelease/index1.html}{http://nesssi.cacr.caltech.edu/DataRelease/index1.html}}, 
and MEarth\footnote{\href{https://lweb.cfa.harvard.edu/MEarth/Welcome.html}{https://lweb.cfa.harvard.edu/MEarth/Welcome.html}} projects, as well as the Mikulski Archive for Space Telescopes (MAST) for Kepler/K2 data\footnote{\href{https://mast.stsci.edu/portal/Mashup/Clients/Mast/Portal.html}{https://mast.stsci.edu/portal/Mashup/Clients/Mast/Portal.html}}.  
 Gaia astrometric data are available from the CDS/VizieR service (\citealt{Gaia2021, Gaia2023}).



\bibliographystyle{mnras}
\bibliography{paper} 




\appendix

\section{Signal-to-Noise and Equivalent-Width Estimation}\label{app:snr_calc}

To facilitate a uniform comparison across instruments, we re-derived the SNR directly from the reduced one-dimensional spectra using a consistent methodology.  
For each exposure, the continuum-normalised flux was examined in two line-free regions adjacent to the Li\,\textsc{i}\,6708\,\AA\ feature (6702–6706\,\AA\ and 6710–6714\,\AA). The per-pixel SNR was defined as the mean flux divided by a robust estimate of the noise, calculated as \(1.4826\times\mathrm{MAD}\), where MAD denotes the median absolute deviation.  
The SNR per resolution element was then obtained as
\begin{equation}
\mathrm{SNR}_{\mathrm{res}} = (\mathrm{SNR}_{\mathrm{pix}})\,\sqrt{N_{\mathrm{pix/res}}},
\end{equation}
where \(N_{\mathrm{pix/res}} = \mathrm{FWHM}/\delta x\), with \(\mathrm{FWHM} = \lambda / R\) representing the instrumental resolution element and \(\delta x\) the median pixel spacing near 6708\,\AA.  

The equivalent widths of the Li\,\textsc{i}\,6708\,\AA\ doublet was measured by direct integration of \(1-F_\lambda\) within a \(\pm0.3\,\AA\) window centred on 6707.8\,\AA, following local continuum normalisation over 6702–6714\,\AA.  
The associated uncertainty, \(\sigma_{\mathrm{EW}}\), was estimated using the relation of \citet{Cayrel1988},
\begin{equation}
\sigma_{\mathrm{EW}} = 1.6\,\frac{\sqrt{\mathrm{FWHM}\,\delta x}}{\mathrm{SNR}},
\end{equation}
and the \(3\sigma\) upper limits reported in Table~\ref{tab:lithium_ew} correspond to \(3\,\sigma_{\mathrm{EW}}\).  

\section{Rotation Period Determination}\label{app:rot_period}

The stellar rotation periods of TOI-736 and TOI-1468 were derived using the Gaussian Process (GP) approach described in \citet{Lalitha2025}, implemented with the \texttt{celerite2} framework.
This method employs a quasi-periodic covariance kernel to model rotationally modulated stellar variability.

\begin{figure}
    \centering
    \includegraphics[width=\linewidth]{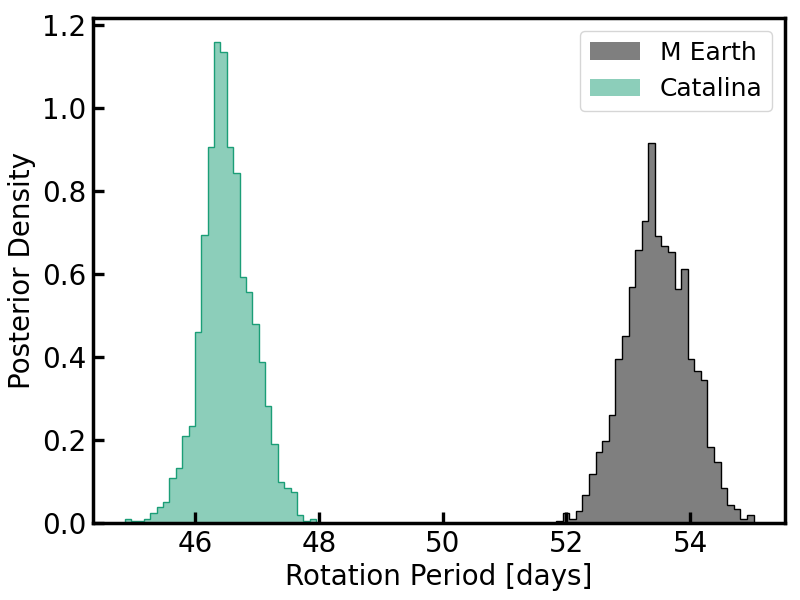}
    \caption{Gaussian process posterior distributions of the rotation period derived from the MEarth and CATALINA photometric light curves of TOI-736.}
    \label{fig:rotation_toi736}
\end{figure}

For TOI-736, we analysed photometric light curves from the MEarth and CATALINA surveys after removing outliers and long-term trends.  The GP analysis (Fig.~\ref{fig:rotation_toi736}) yields rotation periods of $53.4^{+0.6}_{-0.4}$~days (MEarth) and $46.4^{+0.5}_{-0.3}$~days (CATALINA). The two solutions are not formally consistent, and we therefore adopt the range spanned by their combined posterior, 46--57 days, rather than a single merged value. Periods measured in different bandpasses and epochs can differ in the presence of spot evolution or differential rotation, and a forced merge would understate that spread.

\begin{figure}
    \centering
    \includegraphics[width=\linewidth]{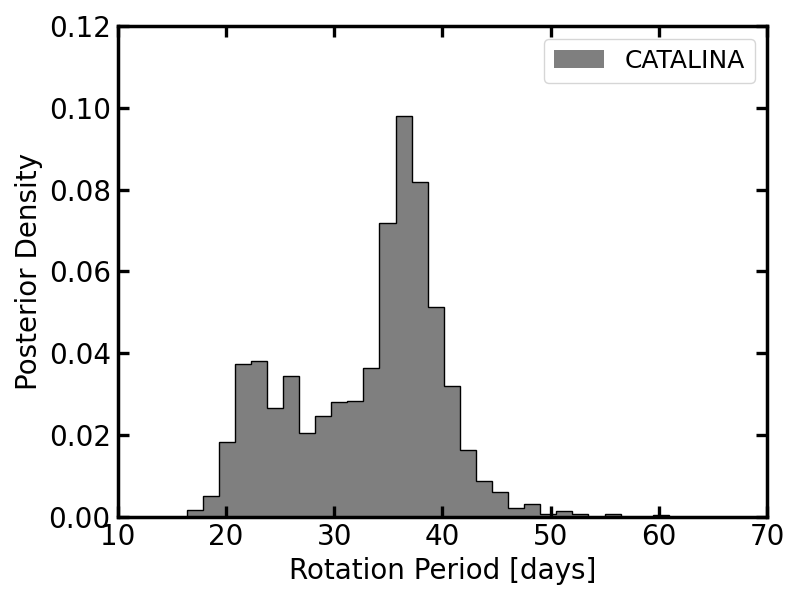}
    \includegraphics[width=\linewidth]{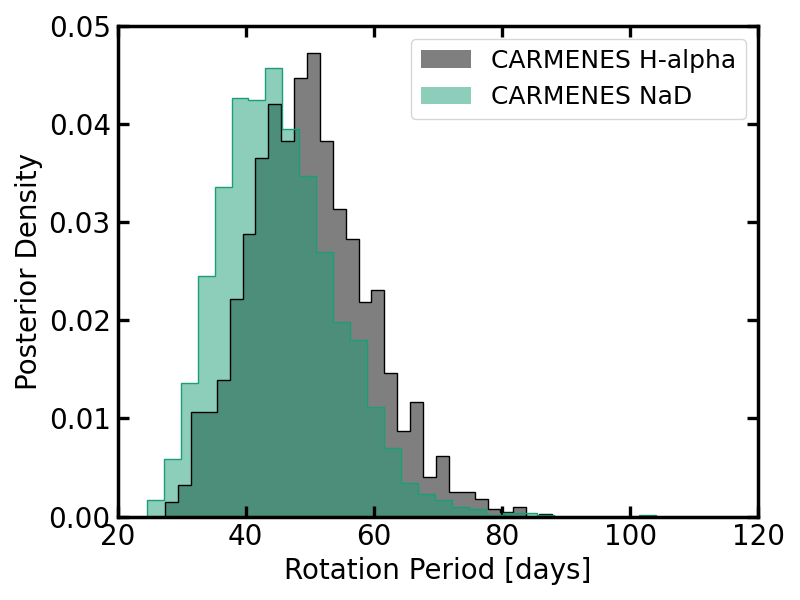}
    \caption{Gaussian process analysis of the rotation period of TOI-1468 from CATALINA photometry (top) and CARMENES spectroscopic activity indices (bottom). 
  }
    \label{fig:rotation_toi1468}
\end{figure}

For TOI-1468, we analysed CATALINA photometry and CARMENES spectroscopic activity indices (Na \textsc{D} and H$\alpha$), shown in Fig.~\ref{fig:rotation_toi1468}. The GP model captures coherent quasi-periodic modulation across all datasets, yielding rotation periods of $37.2^{+1.9}_{-13.1}$~days (CATALINA), $43.4^{+11.3}_{-7.0}$~days (Na \textsc{D}), and $40.9^{+18.6}_{-0.2}$~days (H$\alpha$).
Despite the broad formal posteriors, all three estimates converge within the 37--43~day range, indicating a representative stellar rotation period of $40 \pm 3$~days. Our GP analysis is consistent with the spectroscopic determination of \citet{Chaturvedi2022}, whose value of 41--44 days we adopt, as it is based on a direct spectroscopic time series.

\section{Rotation period and chromospheric activity index relationship}
\label{app:rhk_prot}

To infer a rotation period for TOI-1231 in the absence of a direct time-series determination, we first derived its chromospheric activity index, $\log R'_{\mathrm{HK}}$, from the measured Mount Wilson $S_{\mathrm{MW}}$ value. This calibration follows the prescriptions of \citet{Noyes1984} and \citet{Suarez2015}, which correct for continuum flux variations and subtract the photospheric contribution to the Ca\,\textsc{ii}\,H\&K lines. The relevant relations are reproduced below for completeness.

\begin{equation}
    \log R'_{\mathrm{HK}} = \log_{10}\left(1.34 \times 10^{-4} \, C_{\mathrm{cf}} \, \langle S_{\mathrm{MW}} \rangle - R_{\mathrm{phot}}\right), \label{eqn:rhk}
\end{equation}

where $C_{\mathrm{cf}}$ is a colour-dependent conversion factor that scales the continuum flux to the bolometric luminosity:

\begin{multline}
    \log C_{\mathrm{cf}} = 0.668 - 1.270(B-V) + 0.645(B-V)^2 - 0.443(B-V)^3,
\end{multline}

and $R_{\mathrm{phot}}$ is the photospheric contribution to the Ca\,\textsc{ii} core flux \citep{Hartmann1984}:

\begin{equation}
    R_{\mathrm{phot}} = 1.48 \times 10^{-4} \, e^{-4.3658(B-V)}.
\end{equation}

For TOI-1231 ($B-V = 1.41$), we computed $\log R'_{\mathrm{HK}} = -5.35 \pm 0.07$. This value was then used with the empirical $R'_{\mathrm{HK}}$--$P_{\mathrm{rot}}$ relation to infer the stellar rotation period. To establish this relation, we investigated the correlation between the rotation periods ($P_{\mathrm{rot}}$) and the chromospheric activity index ($R'_{\mathrm{HK}}$) for a sample of M dwarfs, combining data from \citet{Suarez2017}, \citet{Astudillo2017}, and \citet{Shan2024}. We then computed the first derivative of the smoothed distribution to identify regions of significant change in the trend, with particular focus on the point where the slope transition occurs. This threshold was defined as the location where the slope of the curve exhibited the largest change.

The data were divided into two segments based on the identified threshold of P$_{\text{rot,thres}} \approx 17.54$ days. Stars with rotation periods below this threshold were analysed separately from those with periods above it. For each segment, we fit a linear model to the data using least squares fitting.

The fitted linear models for the two regions are characterized by the following parameters:
\begin{multline}
\log(\text{R}_{\text{HK}}) = (0.1075 \pm 0.3097) \cdot \log(\text{P}_{\text{rot}}) \\
+ (-4.4027 \pm 0.2250) \quad \text{(Below Threshold Region)}
\end{multline}

\begin{multline} \label{eq:A2}
\log(\text{R}_{\text{HK}}) = (-1.0059 \pm 0.4344) \cdot \log(\text{P}_{\text{rot}}) \\
+ (-3.3458 \pm 0.7344) \quad \text{(Above Threshold Region)}
\end{multline}

\begin{figure}
\includegraphics[width=0.5\textwidth]{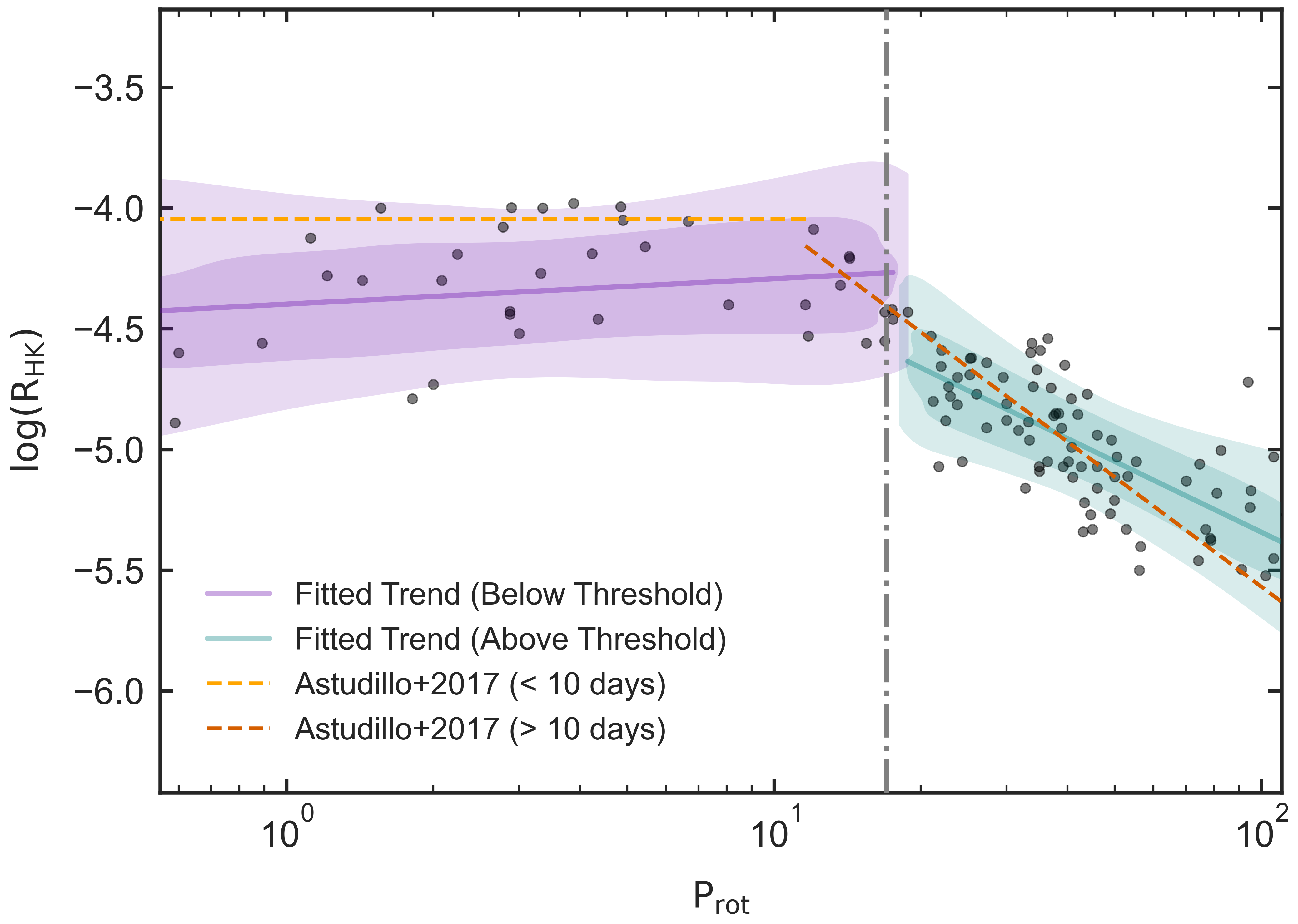}
\caption{Relationship between $\log(R'{_\mathrm{HK}})$ and $\log(P{_\mathrm{rot}})$ for M dwarfs. The plot includes the observed data points, linear fits for regions below and above the threshold rotation period (P$_{\mathrm{rot,thres}} \approx 17.54$ days), and posterior samples from MCMC analysis. The vertical line marks the identified transition point in chromospheric activity behaviour.}
\label{fig:rhk_prot_plot}
\end{figure}

We performed a MCMC sampling of the posterior distributions of the model parameters for both regions. After obtaining the MCMC samples, we calculated the mean and standard deviation of the slope and intercept for each region, providing both the best-fit values and their uncertainties. The fitted models for the two regions (below and above the threshold) were then used to calculate the rotation periods corresponding to specific values of $R_{\text{HK}}$, along with their uncertainties. In Figure~\ref{fig:rhk_prot_plot}, we show the observed data, the fitted trend lines for both regimes, and the posterior contours derived from the MCMC analysis. The dashed lines represent the empirical relations from \citet{Astudillo2017} for the saturated ($P_{\mathrm{rot}} \lesssim 10$~days) and unsaturated ($P_{\mathrm{rot}} \gtrsim 10$~days) regimes, respectively.The vertical dashed line marks the transition between the saturated and unsaturated regimes. For TOI-1231, the measured $\log R'_{\mathrm{HK}}$ places it firmly in the unsaturated regime, corresponding to a rotation period of $98.2 \pm 1.1$~days.

\label{lastpage}
\end{document}